\documentclass[aps,groupeaddress,prd,english,floatfix]{revtex4-1}
\usepackage{epsfig}
\usepackage{amsmath}
\usepackage{slashed}
\usepackage{subfigure,graphicx}
\usepackage{axodraw}
\usepackage{hyperref}

\newcommand{\be}{\begin{equation}}
\newcommand{\ee}{\end{equation}}
\newcommand{\bea}{\begin{eqnarray}}
\newcommand{\eea}{\end{eqnarray}}
\newcommand{\bwt}{\begin{widetext}}
\newcommand{\ewt}{\end{widetext}}

\newcommand{\Tr}{\mathrm{Tr}}

\begin{document}
\title{Probing Radiative Neutrino Mass Generation through Monotop Production}

\author{John N. Ng and Alejandro de la Puente}
\affiliation{Theory Group, TRIUMF, Vancouver BC V6T 2A3, Canada}

\date{\today}

\begin{abstract}

We present a generalization of a model where the right-handed up-type quarks serve as messengers for neutrino mass generation and as a portal for dark matter. Within this framework the Standard Model is extended with a single Majorana neutrino, a coloured electroweak-singlet scalar and a coloured electroweak-triplet scalar. We calculate the relic abundance of dark matter and show that we can match the latest experimental results. Furthermore, the implications from the scattering between nuclei and the dark matter candidate are studied and we implement the latest experimental constraints arising from flavour changing interactions, Higgs production and decay and LHC collider searches for a single jet and jets plus missing energy. In addition, we implement constraints arising from scalar top quark pair-production. We also study the production of a single top in association with missing energy and calculate the sensitivity of the LHC to the top quark semileptonic decay mode with the current $20$ fb$^{-1}$ data set at a centre of mass energy of $\sqrt{s}=8$ TeV. Furthermore, we carry out the analysis to centre of mass energies of $\sqrt{s}=14$ TeV with $30$ and $300$ fb$^{-1}$ of data.

\end{abstract}

\maketitle

\section{Introduction}

We now have ample evidence pointing towards the existence of dark matter~\cite{Bergstrom:2000pn,Bertone:2004pz}. Gravitational phenomena such as velocity dispersion and rotation curves of galaxies suggest the existence of non-luminous matter not composed of the known Standard Model (SM) particles~\cite{Begeman:1991iy,Bradac:2006er}. The most recent data from the Planck collaboration, which builds upon the successful findings of WMAP~\cite{Bennett:2012zja}, estimates a cold dark matter cosmological parameter $\Omega_{\text{DM}}h^{2}=0.1199\pm0.0027$~\cite{Ade:2013zuv}. However, due to the solely gravitational evidence regarding the existence of dark matter, its identity remains unknown. One candidate explanation for dark matter is the existence of a weakly interacting massive particle (WIMP). Models of beyond the SM physics such as the Minimal Supersymmetric Standard Model (MSSM) provide a natural WIMP and reproduce the right abundance in the universe determined by their self-annihilation rate. Furthermore, WIMPs may scatter off nuclei in direct detection experiments, and one can probe the spin dependent~\cite{COUPP,PICASSO,SIMPLE,XENON10,XENON100} and spin independent scattering cross sections~\cite{Ahmed:2010wy,Ahmed:2009zw,Aprile:2012nq,Akerib:2013tjd}. These experiments, together with the relic abundance of dark matter in the universe provide stringent constraints on models beyond the SM with a WIMP candidate.

With the LHC running, scales beyond the electroweak scale are now being probed. The search for new particles and interactions is now underway. In particular, searching for collider signatures that may point at the nature of dark matter has become an active program by both the CMS and ATLAS collaborations. Current dark matter searches are usually carried out by analyzing the possibility that jets are produced in association with a large amount of missing transverse energy (MET)~\cite{ATLAS:2012zim,TheATLAScollaboration:2013fha,CMS:rwa,CMS:2013gea}. Furthermore, final states that do not appear in the SM at tree level have been proposed in order to enhance the sensitivity to dark matter production. One such final state was proposed in~\cite{maltoniMonotop}, where a single top quark is produced in association with MET. This final state, in analogy to its light-quark analogue (monojet), has being named monotop. Monotop production provides a signal that is easier to discriminate than monojet production since the top quark fixes the flavour in the final state. The CDF collaboration carried out a search with $7.7$ fb$^{-1}$ of data at $1.96$ TeV centre of mass energies~\cite{Aaltonen:2012ek} and very recently the CMS collaboration in the hadronic decay mode of the top quark with $19.7$ fb$^{-1}$ of data at $8$ TeV centre of mass energies~\cite{CMS:2014hba}. They were able to set $95\%$ confidence level upper limits on the cross section for the process $pp\to t+\text{MET}$. Models beyond the SM with a monotop signature have been proposed and in particular models where an effective theory approach has been taken have analyzed the significance of a signal in both the hadronic and semi-leptonic decay modes of the top quark with the $7$ and $8$ TeV data sets at the LHC~\cite{maltoniMonotop,wangMonotop,Agram:2013wda}. Top-down approaches have also been studied. In particular, a charged $Z'$ model was proposed to address the non-zero forward-backward top asymmetry at the Tevatron and the null charge asymmetry at the LHC~\cite{Zp_1,Zp_2}. A monotop signal can also arise within a Type II Two-Higgs doublet model supplemented with a SM gauge singlet scalar, identified as the dark matter candidate~\cite{2HDM_1}. The significance of a monotop signal in the last two models has been analyzed in~\cite{Alvarez:2013jqa} and a future search by the LHC can probe the $Z'$ model with the current $20$ fb$^{-1}$ of data at $8$ TeV centre of mass energies. Another model implements a coloured electroweak-singlet scalar and three Majorana fermions, uncharged under the SM gauge symmetry, to address the non-zero forward-backward top asymmetry through on shell production of the coloured scalar~\cite{Kumar:2013jgb}. Furthermore, supersymmetric models with $R$-parity violation can lead to singly produced top quarks in association with a long-lived neutralino~\cite{SUSY_1,SUSY_2,SUSY_3}. In this work we revisit the model introduced in~\cite{ModelMain} where, without extending the SM gauge symmetry, we incorporate a Majorana neutrino coupling to up-type right-handed quarks through a coloured electroweak-singlet scalar. The simplest and most economical case studied there used only the top and charm quarks. In addition, we add a coloured electroweak-triplet scalar coupling to left-handed leptons and left-handed quarks. Our model belongs to a family of models where the dark matter relic abundance is induced by the exchange of a $t$-channel scalar mediator and where the dark matter nucleon scattering cross section is through the $t$-channel exchange of the mediator~\cite{Bai,Chang,An,DiFranzo:2013vra,Garny:2014waa}. Furthermore, within our framework active neutrino masses are generated radiatively and can be understood in a generalized seesaw framework. The active neutrino masses are given by the effective $\frac{g_{eff}}{\Lambda}(LH)^{2}$ operator where the seesaw scale is $\Lambda$ and $g_{eff}$ is some effective coupling. For our model, $\Lambda \simeq O(\mathrm {TeV})$, is
determined by the mass of the heavy coloured scalar. The coupling $g_{eff}$ is a product of 3 loop-factors, several Yukawa and scalar couplings of which the top Yukawa coupling is the largest. Hence, the top quark plays a very special role.  In~\cite{ModelMain}, the right-handed up quark was not coupled to the Majorana neutrino and the coloured electroweak-singlet and thus we were able to easily evade most dark matter direct detection constraints and current monojet and jets~+~MET constraints at the LHC. Similarly within that framework, new collider signatures that can be probed with current and future data, such as monotops, were also absent. Making the up quark play an active role not only completes the model but also provides us with a clear monotop signature that would have been suppressed in the previous simplified model. It would be interesting to witness future analyses by the CMS and ATLAS collaborations  that can probe models such as ours that predict the production of a singlet top quark in association with a dark matter particle. In our case, this could be evidence of the underlying mechanism that bestows neutrinos with mass.

The summary of our study is as follows: In Section~\ref{sec:model} we review the model and in Section~\ref{sec:DarkMatter} the mechanism that leads to the relic abundance of dark matter. Furthermore, we analyze the implications from the scattering between nuclei and our dark matter candidate. In Section~\ref{sec:neutmass} we review the mechanism that radiatively generates active neutrino masses. In Section~\ref{sec:constraints} we analyze all of the experimental constraints sensitive to our framework and in Section~\ref{sec:results} we summarize our results and present the allowed regions of parameter space within different model scenarios. In Section~\ref{sec:monotop} we analyze the monotop signal that arises in our framework and discuss the sensitivity of the LHC to the semileptonic decay modes with the current $20$ fb$^{-1}$ data set at $8$ TeV centre of mass energies. We then carry out an analysis of the future sensitivity at $14$ TeV LHC and conclude in Section~\ref{sec:discussion}.

\section{The Basic Model}\label{sec:model}

Within this framework, without extending the SM gauge symmetry, we incorporate a Majorana neutrino, $N_{R}$, that couples to right-handed up-type quarks through a coloured electroweak-singlet scalar, $\psi$ transforming as a ${\bf (3,1,2/3)}$ under the SM gauge group. We also introduce couplings between left-handed leptons and left-handed quarks through a coloured electroweak-triplet, $\chi$
\begin{equation}
\chi=\begin{pmatrix}
\chi_{2}/\sqrt{2} & \chi_{1} \\
\chi_{3} & -\chi_{2}/\sqrt{2}
\end{pmatrix},
\end{equation}
transforming as a ${\bf (3,3,-1/3)}$ under the SM gauge group. The new operators are parametrized by the following Lagrangian:
\begin{eqnarray}
\label{eq:lag}
-{\cal L}_{BSM}&=&\sum_i y_{\psi}^i\bar{u^i}P_{L}N^{c}\psi+\sum _{\ell , i} \left\{ \lambda_\ell ^i\left[\bar{u^i} P_{R}\left(\chi_{1}\nu^{c}_\ell +\frac{\chi_{2}}{\sqrt{2}}\ell^{c} \right)+\bar{d^i} P_{R}\left(\chi_{3}\ell^{c} -\frac{\chi_{2}}{\sqrt{2}}\nu^{c}_\ell \right)\right]\right\}+~\text{hc},\nonumber \\
~~
\end{eqnarray}
where $l=e,\mu,\tau$ and $i=1,2,3$ is the quark family index. Also $P_L$ and $P_R$ are the left- and right-handed projection operators. The coupling $y^{i}_{\psi}$ denotes the strength of the interaction between $N_{R}$ and $u^{i}_{R}$ via $\psi$, while $\lambda^{i}_{l}$ denotes the strength between the quark doublets $(u^{i},d^{i})_{L}$ and lepton $(\nu,l)_{L}$ via $\chi$.

Within this model we also incorporate an arbitrary mass parameter for the Majorana neutrino, $M_{N_{R}}$, in the Lagrangian and introduce a $Z_{2}$ parity, denoted dark parity (DP), into the model. We let $N_{R}$ and $\psi$ be odd under it, with all other fields transforming evenly under it. This assignment is useful since it will allow us to identify $N_{R}$ with a dark matter candidate for $M_{N_{R}}< m_{\psi}$. Throughout this work we take $m_{\psi}>M_{N_{R}}$ and parametrize the mass term for $N_{R}$ with the usual Majorana mass $\frac{1}{2}\, M_N\,\bar{N^c_{R}}\,N_R$.

 We also introduce the most general gauge and $Z_{2}$ symmetric scalar Lagrangian:
\begin{eqnarray}
\label{eq:spot}
V(H,\psi,\chi)&=& -\mu^2 H^{\dagger} H +\frac{\lambda}{4!}(H^{\dagger} H)^2 +m_\chi^2 \Tr (\chi^{\dagger}\chi)+\lambda_\chi (\Tr \chi^{\dagger} \chi)^2 \nonumber \\
&+& m_\psi ^2 \psi^{\dagger}\psi + \lambda_\psi (\psi^{\dagger}\psi)^2+\kappa_1 H^{\dagger} H \Tr \chi^{\dagger} \chi +\kappa_2 H^{\dagger} \chi^{\dagger} \chi H  \nonumber \\
&+&\kappa_3 H^{\dagger} H \psi^{\dagger} \psi+\rho_1 (\Tr \chi^{\dagger} \chi)\psi^{\dagger} \psi,
\end{eqnarray}
where $H$ is the SM Higgs field. We emphasize that in order to avoid a colour breaking vacuum, $m_\chi^2$ and $m_\psi^2$ must be positive. Furthermore, one can see that in this framework, the $Z_{2}$ dark parity remains exact after electroweak symmetry breaking.

\section{Dark Matter}\label{sec:DarkMatter}
\subsection{Relic Abundance}\label{subs:RelicAbundance}

\begin{figure}[ht]
\begin{center}
\begin{picture}(180,120)(-80,-80)
\Text(-45,42)[!]{$N_{R}~(p_{1})$}
\Line(-35,35)(0,0)
\ArrowLine(0,0)(35,35)
\Text(44,42)[!]{$t,c,u~(p_{3})$}
\Vertex(0,0){2}
\Text(20,-25)[!]{$\psi$}
\DashLine(0,-50)(0,0){5}
\Vertex(0,-50){2}
\Text(-44,-95)[!]{${N_{R}~(p_{2})}$}
\Line(0,-50)(-35,-85)
\ArrowLine(35,-85)(0,-50)
\Text(44,-95)[!]{$\overline{t},\overline{c},\overline{u}~(p_{4})$}
\end{picture}
\end{center}
\caption{\small Interaction channels that lead to a reduction in the relic abundance of $N_{R}$: Channels consist of annihilation into $u_{i}\bar{u}_{j}$, $i,j=u,c,t$. A similar $u$-channel graph is not shown. \label{fig:AnnChan}}.
\bigskip
\bigskip
\end{figure}
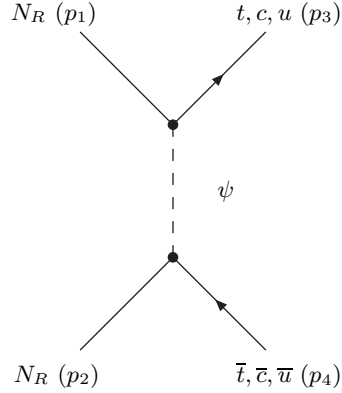

The existence of an unbroken $Z_{2}$ symmetry stabilizes $N_{R}$ and the nature of the Lagrangian introduced in Equation~(\ref{eq:lag}) yields a mechanism for its relic abundance. This mechanism is the $N_R$ pair annihilation through a $t$- and $u$-channel exchange of the new coloured electroweak-singlet, $\psi$. The $u$-channel is available because $N_R$ has similar properties as a Majorana fermion. These diagrams are depicted in Figure~\ref{fig:AnnChan}.

 The present day relic abundance is given by~\cite{Kolb:1990vq}
\begin{equation}
\Omega_{DM}h^2\approx 1.65\times 10^{-10}\left(\frac{\mathrm{GeV}^{-2}}{\left<\sigma v\right>}\right)\log\left(0.038g\frac{mM_{Pl}\left<\sigma v\right>}{g_{*}^{1/2}}\right),
\end{equation}
where we have dropped the temperature dependence in the logarithm since it is not an important factor in the region we are studying. The most important controlling factor above is $\left<\sigma v\right>$ and the correct relic abundance can be achieved with a value of $\left<\sigma v\right> \simeq 3\times 10^{-26}~\text{cm}^3/\text{s}$.

The thermalized cross section at temperature $T$ can be calculated from the annihilation cross section of our dark matter candidate, $N_{R}$. The annihilation channels are depicted in Figure~\ref{fig:AnnChan}. The thermalized cross section is given by
\begin{equation}
\left<\sigma_{N_{R}N_{R}} v\right>=\int^{\infty}_{4M^{2}_{N_{R}}}ds\frac{(s-4M^{2}_{N_{R}})s^{1/2}K_{1}(s^{1/2}/T)}{8M_{N_{R}}^4 TK_{2}^{2}(M_{N_{R}}/T)}\sigma(s),
\end{equation}
where $\sigma(s)$ is the annihilation cross section of the $2\to2$ annihilation process
\begin{equation}
\frac{d\sigma (s)}{d\Omega}=\frac{|M|^{2}}{64\pi^{2}s}\frac{|\vec{p}_{3}|}{|\vec{p}_{1}|},
\end{equation}
and $K_{1}(z),K_{2}(z)$ are Modified Bessel functions of the first and second kind respectively. This procedure requires elaborate
numerical computations. Below we outline a good approximate calculation of $\sigma v$ which we check against the numerical
calculations.

In our analysis we use the Boltzmann distribution to calculate the thermal averaged cross section. This allows us to consider a low velocity expansion of the annihilation cross section:
\begin{equation}
\sigma v_{rel}\approx a_{rel}+b_{rel}v^{2}_{rel}+...+,
\end{equation}
where $v_{rel}$ is related to the center of mass velocity of the annihilating particle by $v_{CM}=v_{rel}/2$. We also make use of a technique studied in~\cite{Wells:1994qy} which uses the fact that Mandelstam variables can be expanded in the following way
\begin{eqnarray}
s&=&s_{0}+s_{2}v^{2}_{CM}+...+ \nonumber \\
t&=&t_{0}+t_{1}\cos\theta v_{CM}+t_{2}v^{2}_{CM}+...+
\end{eqnarray}

With this in mind, and using the notations of ~\cite{Wells:1994qy}, the differential cross section can be written as
\begin{equation}
v_{cm}\frac{d\sigma (s)}{d\Omega}=\frac{J(s,t)}{4\pi}K(s)\sqrt{1-v^{2}_{CM}},
\end{equation}
where $J(s,t)=|M|^{2}$ and $K(s)=|\vec{p}_{3}|/16\pi ms$ can both be expanded about $s=s_{0}$ and $t=t_{0}$ to obtain
\begin{eqnarray}
a_{rel}&=&2J_{0}K_{0} \nonumber \\
b_{rel}&=&\frac{1}{2}J_{0}K_{2}-\frac{1}{4}J_{0}K_{0}+\frac{1}{2}J_{2}K_{0}.
\end{eqnarray}
For annihilation only into light quarks, the thermalized cross section is $p$-wave suppressed and it is given by
\begin{equation}
\left<\sigma_{N_{R}N_{R}} v\right>\approx v^{2}_{rel}\left[(y^{u}_{\psi})^{4}+(y^{c}_{\psi})^{4}\right]\frac{m^{2}_{N_{R}}(m^{4}_{N_{R}}+m^{4}_{\psi})}{16\pi (m^{2}_{N_{R}}+m^{2}_{\psi})^{4}},
\end{equation}
while for annihilation mainly into $t\bar{t}$, small $y^{u,c}_{\psi}$, the thermalized cross section is dominated by the $s$-wave:
\begin{eqnarray}
\left<\sigma_{N_{R}N_{R}} v\right>&\approx&\frac{3 m_t^2 (y^t_{\psi })^2}{128 \pi  M_{N_R}^4}\left(\frac{4(y^{t}_{\psi})^{2}M^{3}_{N_{R}}\sqrt{(M_{N_{R}}-m_{t})(M_{N_{R}}+m_{t})}}{(M^{2}_{N_{R}}-m^{2}_{t}+m^{2}_{\psi})^{2}}\right. \nonumber \\
&-&\left.\frac{[(y^{u}_{\psi})^{2}+(y^{c}_{\psi})^{2}](4M^{2}_{N_{R}}-m^{2}_{t})^{2}}{2(2M^{2}_{N_{R}}-m^{2}_{t}+2m^{2}_{\psi})^{2}}\right). \nonumber \\
\end{eqnarray}

\begin{figure}[ht]\centering
\subfigure{\includegraphics[width=3.0in]{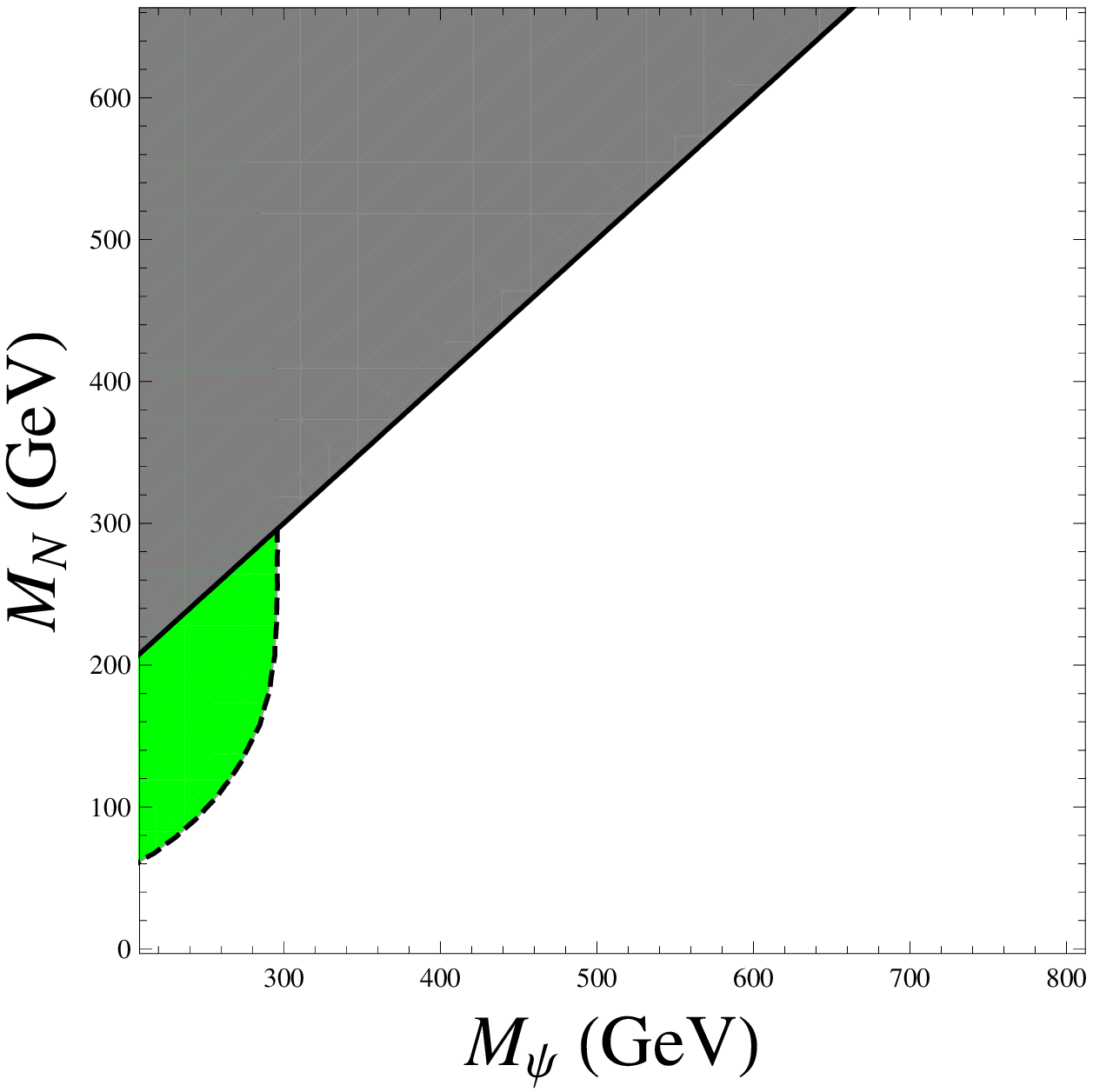}\label{fig:DM_G2}}~
\subfigure{\includegraphics[width=3.0in]{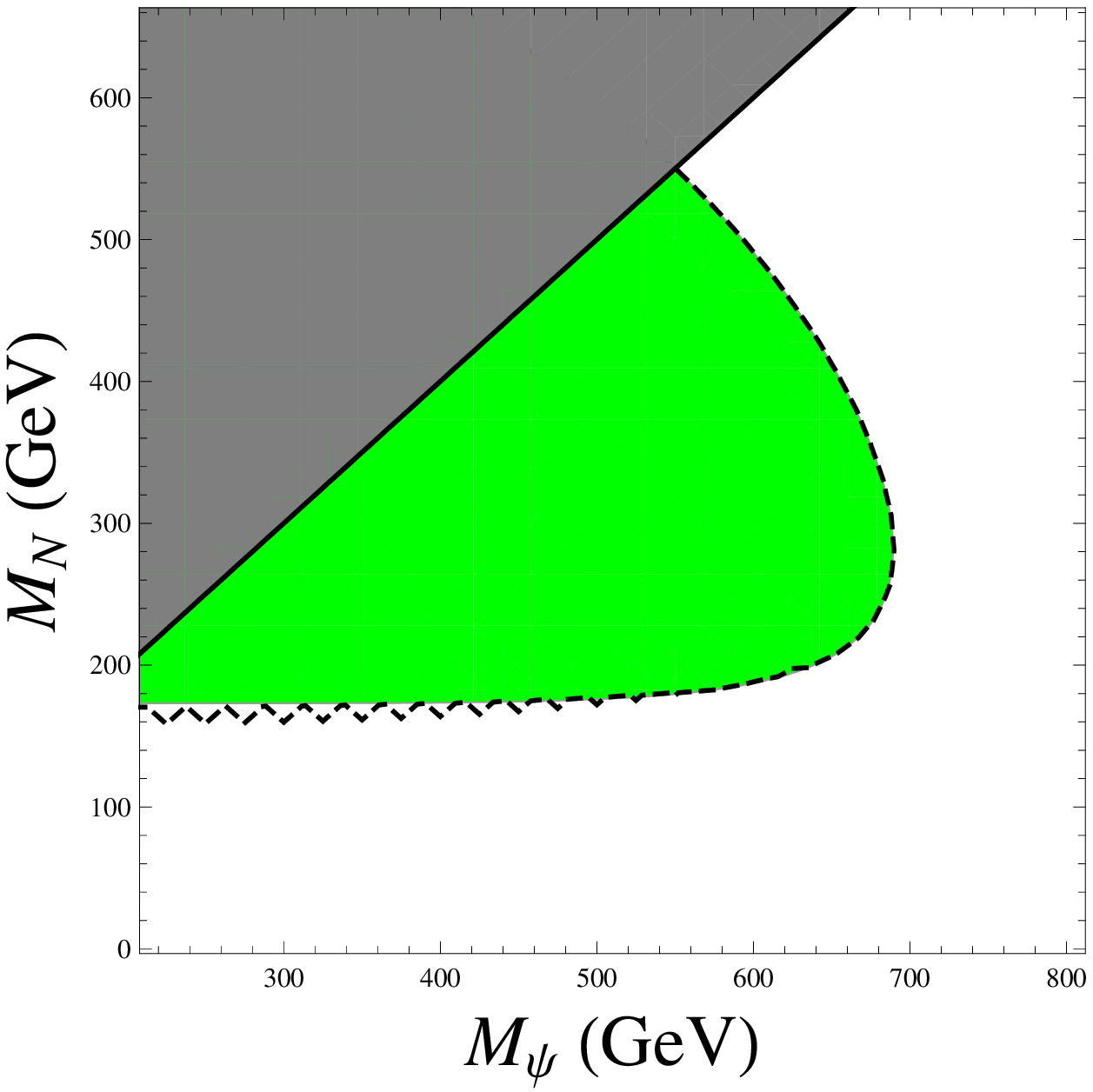}\label{fig:DM_G1}}
\caption{\small
Thermalized cross section in the $m_{\psi}-M_{N_{R}}$ plane for $y^{t,u}_{\psi}=0$, $y^{c}_{\psi}=1$ on the left and $y^{c,u}_{\psi}=0$, $y^{t}_{\psi}=1$ on the right. The green region corresponds to $\left<\sigma v\right> >3\times10^{-26}\text{cm}^{3}/\text{s}$ while the dashed black line to $\left<\sigma v\right> =3\times10^{-26}\text{cm}^{3}/\text{s}$.
}
\end{figure}
With the above method, our calculations were in good agreement with the relic abundance generated by MicrOMEGAs~\cite{Belanger:2010gh} with model files generated with FeynRules~\cite{Feynr}. We would also like to emphasize that near the degenerate region, $M_{N_{R}}\approx m_{\psi}$, co-annihilation effects become important. The co-annihilation channels that contribute to the relic abundance calculation are $N_{R}\psi^{\dagger}\to u/c/t, g$ as well as the $\psi\psi^{\dagger}$ annihilation channels. These channels tend to make the annihilation process more efficient requiring lower values of $y^{t,c,u}_{\psi}$ and become important if the mass difference, $\delta m=m_{\psi}-M_{N_{R}}$, is small compared to the freeze-out temperature of the Majorana neutrino~\cite{Griest:1990kh}.  Within our framework, the region consistent with the cold dark matter cosmological parameter sits away from this region and thus co-annihilation effects can be safely neglected. The results for $y^{t,u}_{\psi}=0$ and $y^{c}_{\psi}=1$ are shown in Figure~\ref{fig:DM_G2}. The dashed line corresponds to an annihilation cross section of $\left<\sigma v\right>=3\times10^{-26}$ cm$^{3}$/s, the grey region corresponds to $m_{\psi}<M_{N_{R}}$ and the green region to an annihilation cross section into a pair of charm quarks greater than $3\times10^{-26}$ cm$^{3}$/s . For $y^{t}_{\psi}=1$ and $y^{c,u}_{\psi}=0$ the annihilation will be into $t\bar{t}$ and the $s$ wave contribution dominates. We can see this in Figure~\ref{fig:DM_G1} where the green region corresponds to $s$-wave contributions greater than $3\times10^{-26}$ cm$^{3}$/s.

\subsection{Direct Detection}\label{subs:DirectDetect}
In our framework, the dark matter candidate is a Majorana fermion with chiral symmetric interactions that lead to a non-relativistic WIMP-nucleon scattering cross section dominated by spin-dependent interactions~\cite{Agrawal}. The scattering is dominated by $t$- and $u$-channel exchanges of the coloured electroweak-singlet, $\psi$. The amplitude in terms of the quark-current expectation values is given by
\begin{equation}
{\cal M}=\frac{1}{(m^{2}_{\psi}-M^{2}_{N_{R})}}\bar{u}_{N_{R}}\gamma^{\mu}\gamma^{5}u_{N_{R}}\left<\bar{q}\gamma_{\mu}\gamma^{5}q\right>,
\end{equation}
and the cross section by
\begin{equation}
\sigma^{N}_{SD}=\frac{3(y^{u}_{\psi})^{4}\Delta^{2}_{N}}{64\pi[(m^{2}_{\psi}-M^{2}_{N_{R}})^{2}+\Gamma^{2}_{\psi}m^{2}_{\psi}]}\frac{M_{N_{R}}m_{N}}{M_{N_{R}}+m_{N}}\label{eq:SDcc},
\end{equation}
where $\Delta_{p(n)}=0.78~(-0.48)$ is the spin fraction of the proton (neutron) carried by the $u$-quark~\cite{Mallot}, $m_{N}$ the mass of the nucleon, and $\Gamma_{\psi}$ is the full width of the scalar mediator. The implications of the coupling between the $u$-quark and a Majorana fermion has recently been studied in~\cite{Bai,Chang,An}. However, in these works, the coloured electroweak-singlet couples only to the $up$-quark unlike our framework where it is free to couple to all three generations. Furthermore, the authors in~\cite{An} show that a spin-independent signal can be generated by the following effective operators:
\begin{eqnarray}
O_{1}&=&\frac{\alpha_{S}}{4\pi}G^{a\mu\nu}G^{a}_{\mu\nu}N^{2}_{R} \nonumber \\
O_{2}&=&m_{q}\bar{q}qN^{2}_{R},
\end{eqnarray}
where $G^{a}_{\mu\nu}$ is the gluon field tensor and $\alpha_{S}$ is the strong coupling constant; but this signal is suppressed compared to the spin-dependent contribution given in Equation~\ref{eq:SDcc}. Therefore, for a coupling to protons we compare our prediction to limits set by SIMPLE, COUPP and PICASSO~\cite{SIMPLE,COUPP,PICASSO} which set the most stringent constraints to date and limits from XENON10~\cite{XENON10} for the case where the Majorana neutrino couples to neutrons. The XENON100 results are now available~\cite{XENON100} and we note that they are an order of magnitude stronger than those from XENON10 for dark matter masses above $10$ GeV.
\begin{figure}[ht]\centering
\subfigure{\includegraphics[width=3.0in]{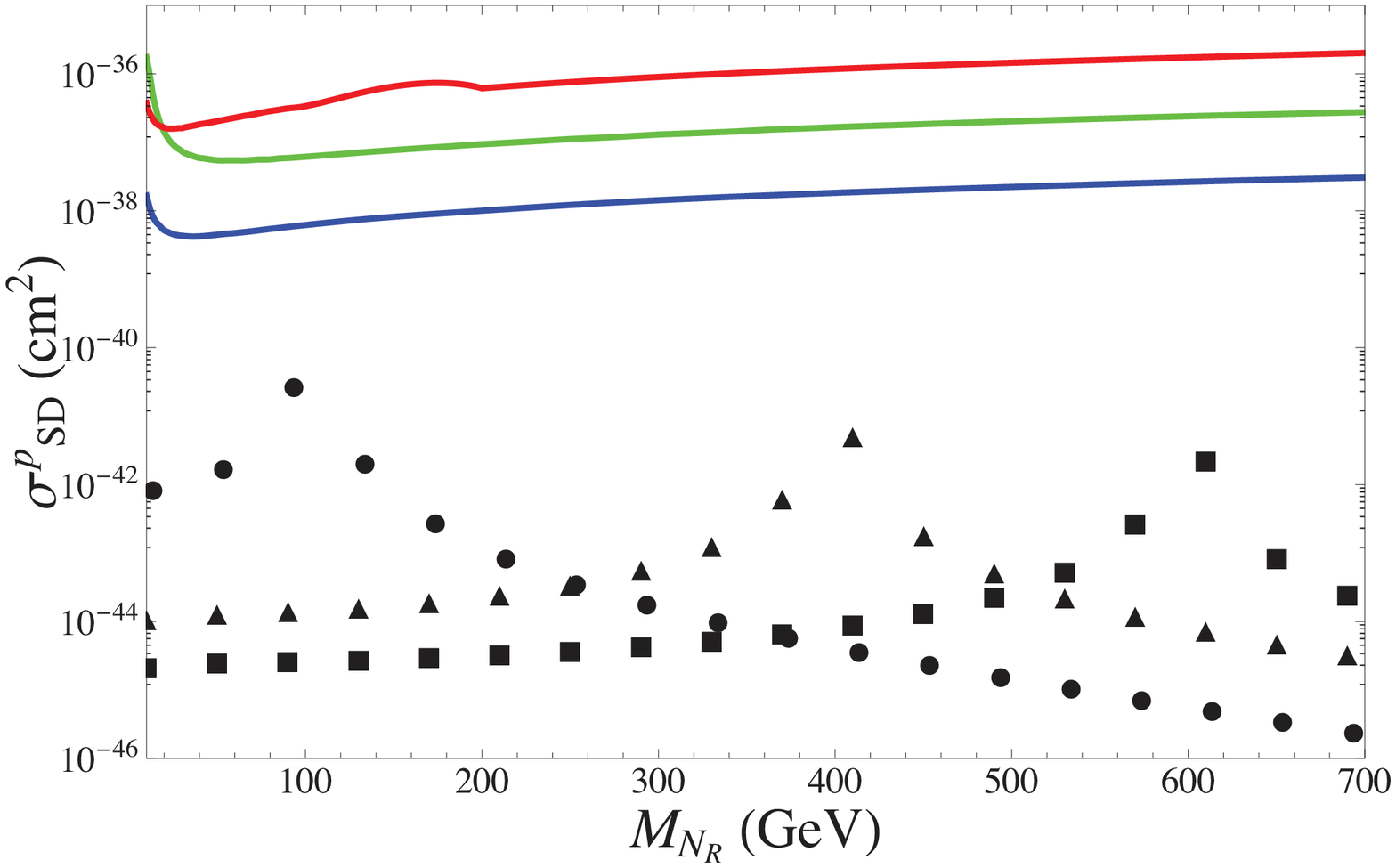}\label{fig:DDproton_G1}}~
\subfigure{\includegraphics[width=3.0in]{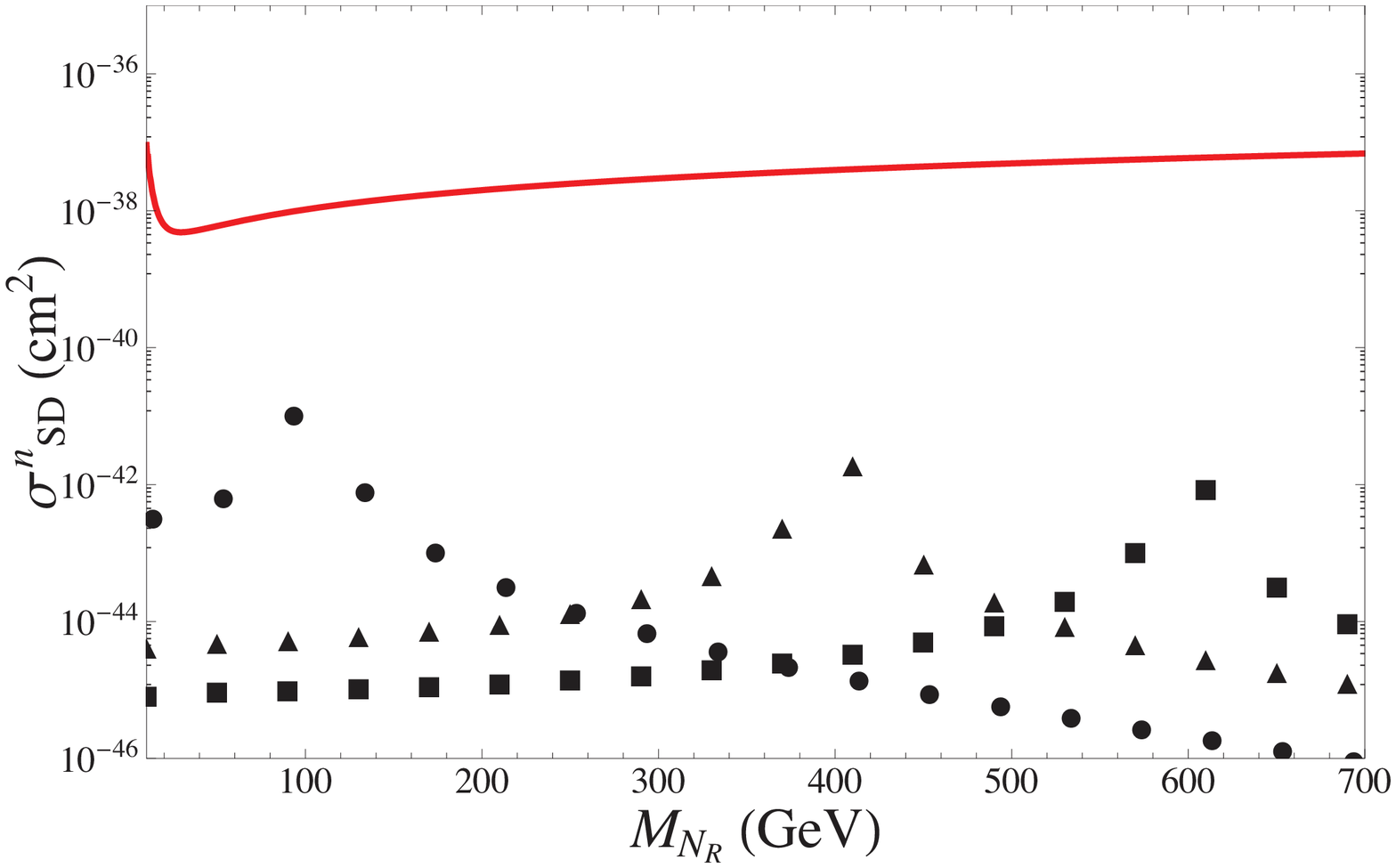}\label{fig:DDneutron_G1}}
\caption{\small Spin dependent cross section as a function of the Majorana neutrino mass, $M_{N_{R}}$ for $y^{u}_{\psi}=0.1$. For coupling to protons (left), limits from  COUPP~\cite{COUPP}, PICASSO~\cite{PICASSO}, and SIMPLE~\cite{SIMPLE} are shown in green, red and blue respectively while the spin dependent cross section for scalar mediator masses of $m_{\psi}=100,400,600$ GeV are depicted by the solid disks, triangles and squares respectively. For coupling to neutrons (right), limits from XENON10~\cite{XENON10} are shown in red.}
\end{figure}

\begin{figure}[ht]\centering
\subfigure{\includegraphics[width=3.0in]{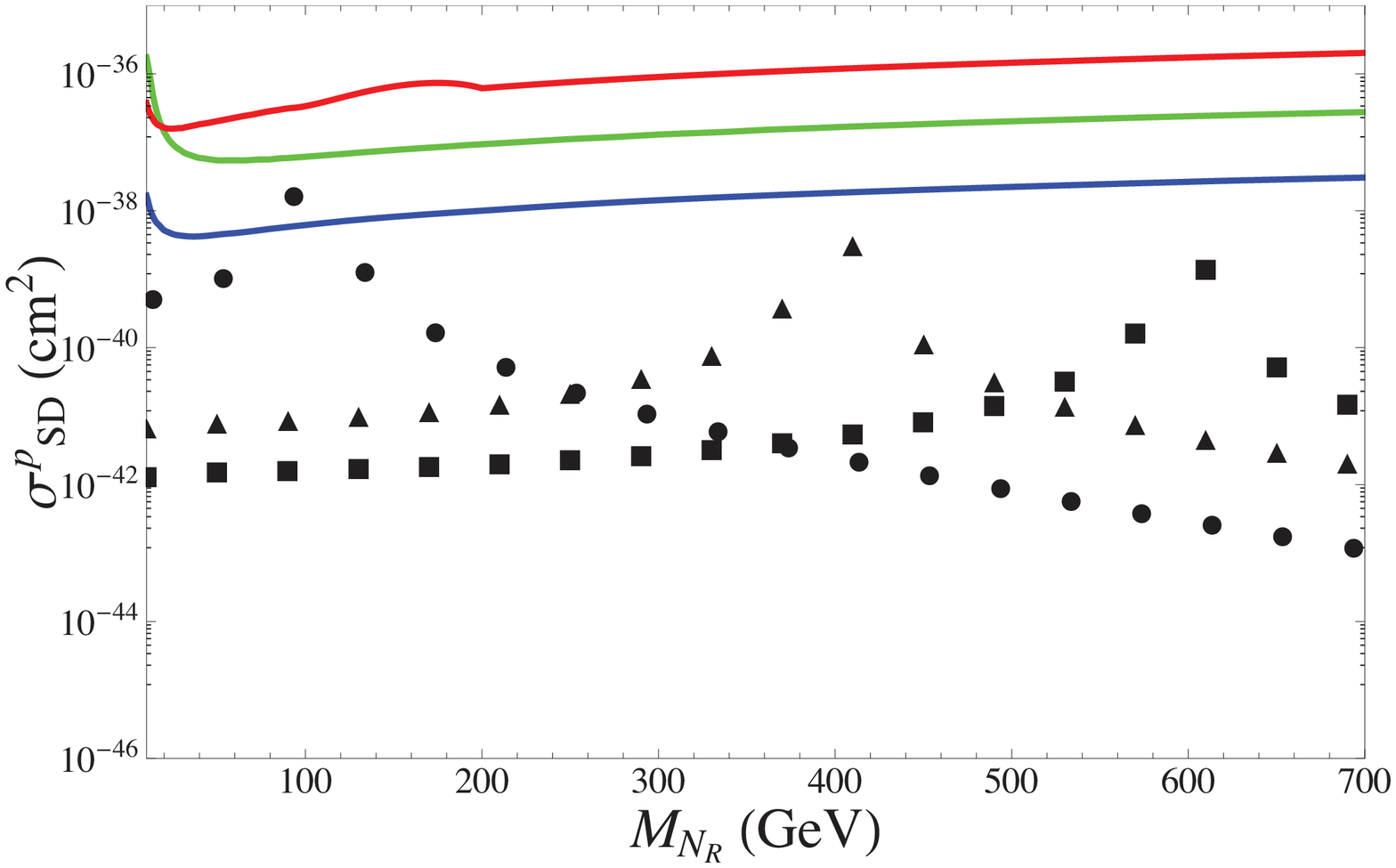}\label{fig:DDproton_G2}}~
\subfigure{\includegraphics[width=3.0in]{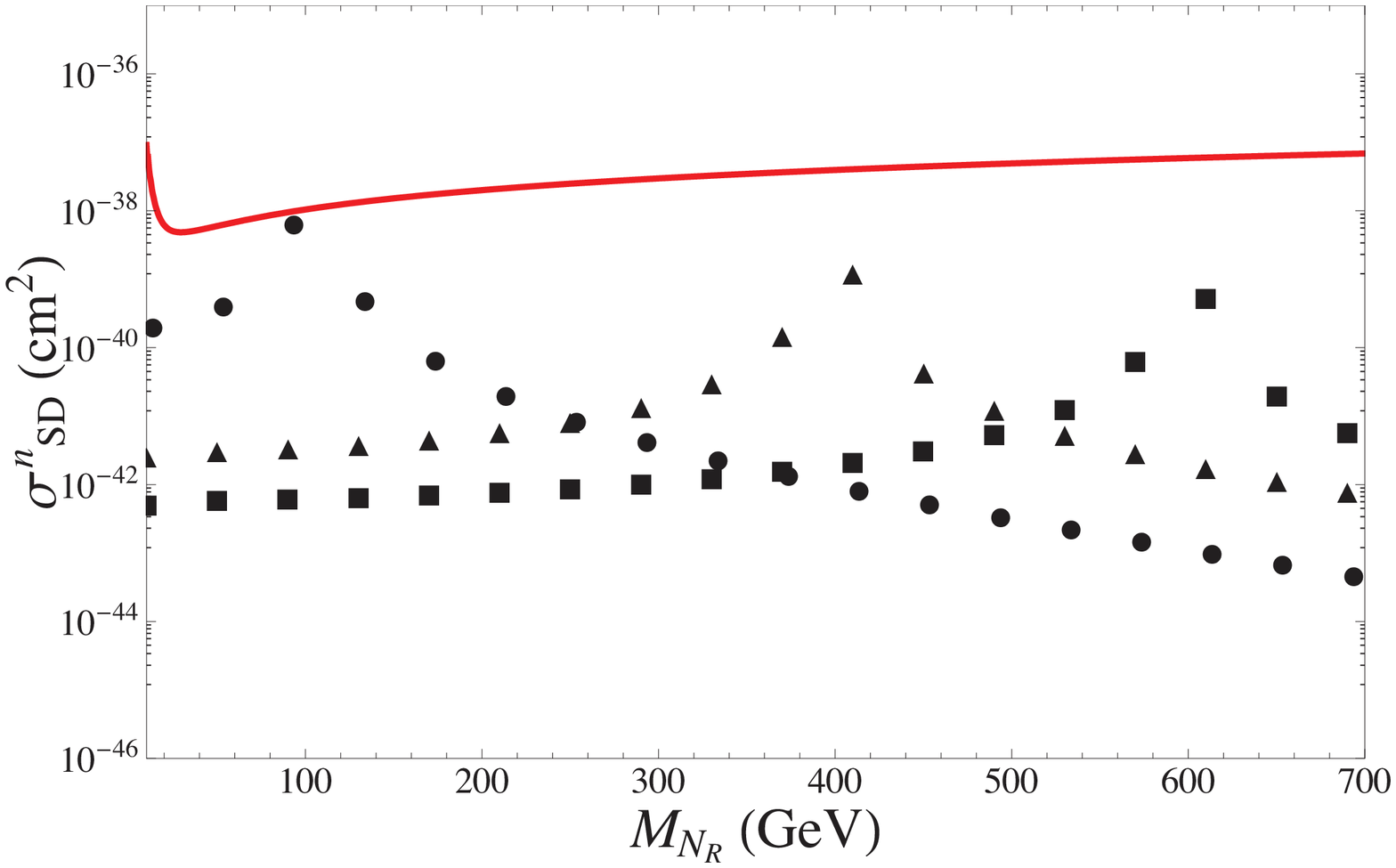}\label{fig:DDneutron_G2}}
\caption{\small Same as in Figure~\ref{fig:DDproton_G1} and~\ref{fig:DDneutron_G1} but with $y^{u}_{\psi}=0.5$ }
\end{figure}

In Figures~\ref{fig:DDproton_G1} and~\ref{fig:DDneutron_G1} we show the spin dependent cross section as a function of the Majorana neutrino mass for coupling to the proton and the neutron respectively with $y^{u}_{\psi}=0.1$. We have made use of DMTools~\cite{DMtools} to plot the limits. In the figures we show the spin dependent cross section for three scalar mediator masses. The solid circles correspond to $m_{\psi}=100$ GeV, while the solid triangles and squares correspond to $m_{\psi}=400,600$ GeV respectively. It is evident from the figures that the cross section is enhanced for $M_{N_{R}}\approx m_{\psi}$, but the corresponding coupling is too small to constrain the model. In Figures~\ref{fig:DDproton_G2} and~\ref{fig:DDneutron_G2} we show the spin-dependent cross section with $y^{u}_{\psi}=0.5$. Here we see that for small coloured electroweak-singlet mediator with mass~$\sim100$ GeV, the region where $m_{\psi}\sim M_{N_{R}}$ is excluded in both proton and neutron scattering. Therefore, direct detection constraints rule out well tuned combinations of couplings and masses as expected from Equation~\ref{eq:SDcc}; and one may want to stay away from the resonant regions for $y^{u}_{\psi}\to1$.

\section{Radiative Neutrino Mass generation}\label{sec:neutmass}

\begin{figure}[ht]\centering
\includegraphics[width=3.0in]{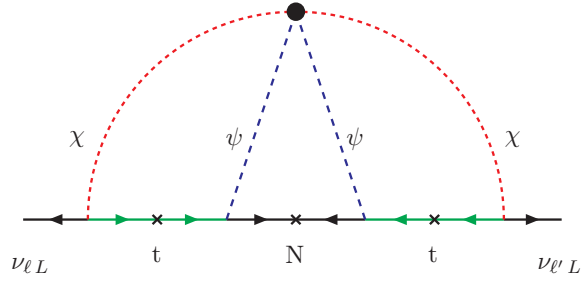}
\caption{\small 3-loop generation of a Majorana mass for active neutrinos from the t-quark. The crosses on the fermion lines indicate mass insertions. The blob at the top of the diagram indicates an effective $\psi\psi\chi\chi$ vertex. Similar diagrams from the $u$- and $c$-quarks will also play a role.}\label{fig:numass}
\end{figure}
In~\cite{ModelMain} it is mentioned that the unbroken DP can be used to forbid Dirac neutrino mass terms for the active neutrinos, $\nu_{i}$. Therefore, within this framework the usual seesaw mechanism is not operative. However, we showed that the Lagrangian of Equation (\ref{eq:lag}) had enough structure to radiatively generate masses for $\nu_{i}$ via the exchange of the exotic coloured scalars. In particular, it had the novel feature of using the right-handed up-type quarks as a portal to communicate with the dark sector. Furthermore, the up-type quarks served as messengers to radiatively generate Majorana masses for the active neutrinos; with the lowest order diagram for neutrino mass generation at three-loops. Models where neutrino masses are generated at the three-loop level have been studied in the past. In particular, the model proposed by Krauss, Nasri, and Trodden (KNT)~\cite{Krauss:2002px} extends the SM with new scalars and a right handed neutrino playing the role of the dark matter. More recently a variation of the KNT model was introduced and incorporates a fermion triplet to generate neutrino masses at the three-loop level~\cite{Ahriche:2014cda}. However, unlike our framework, the new fields are not charged under colour and quarks do not play and active role in the generation of neutrino masses. Within our framework, the three-loop diagram is due to exchanges of both $\psi$ and $\chi$ fields. The mechanism is depicted in Figure \ref{fig:numass}. However within this framework, the loop is closed through an effective $\psi\psi\chi_{1}\chi_{1}$ vertex parametrized by an effective coupling, $\rho$. This vertex is both hypercharge and weak isospin changing. One way of generating this vertex is by introducing a coloured electroweak-triplet scalar, $\omega$
\begin{equation}
\omega=\begin{pmatrix}
\omega_{2}/\sqrt{2} & \omega_{1} \\
\omega_{3} & -\omega_{2}/\sqrt{2}
\end{pmatrix},
\end{equation}
with no tree-level coupling to fermions. The field $\omega$ transforms as a ${\bf (3,3,2/3)}$  under the SM gauge group and it is even under $Z_{2}$. We add to the scalar potential, Equation~(\ref{eq:spot}), the terms involving $\omega$
\begin{eqnarray}
V(H,\psi,\chi,\omega)&=&m_\omega^2 \omega^\dagger\omega +\lambda_\omega (\text{Tr} \omega^\dagger \omega)^2 +\kappa_{4}H^{\dagger}H\text{Tr}\omega^{\dagger}\omega+\kappa_{5}H^{\dagger}\omega^{\dagger}\omega H \nonumber \\
&+&\rho_{2}\left(\text{Tr}\omega^{\dagger}\omega\right)\psi^{\dagger}\psi+ \rho_{3}\text{Tr}\left(\omega^{\dagger}\psi\omega^{\dagger}\psi\right)+\alpha \text{Tr} H^{T}i\sigma_{2}\chi\omega^{\dagger}H+\tilde{V}(\chi,\omega)+\text{h.c},
\end{eqnarray}
where $\tilde{V}(\chi,\omega)$ parametrizes all renormalizable quartic couplings between $\chi$ and $\omega$. Electroweak symmetry breaking leads to $\chi-\omega$ mixing given approximately by $\theta_{\chi-\omega}\sim\alpha v^{2}/(m_{\omega}^2-m_{\chi}^2)$ with $v=\left<H\right>=174$ GeV. Barring accidental degeneracy between $\chi$ and $\omega$ this mixing is expected to be small. Thus, the $\rho_{3}$ coupling in the scalar potential yields a four scalar coupling involving $\psi\psi\omega^{\dagger}_{2}\omega^{\dagger}_{2}$ and an effective vertex between $\psi\psi\chi_{1}\chi_{1}$ arises and it is given by
\begin{equation}
\rho=\rho_{3}\cdot\theta_{\chi-\omega}.
\end{equation}
Since $\omega$ has no couplings to fermions is has less interesting phenomenology than $\chi$ although its mass is of order $m_\chi$. Furthermore, the effective $\rho$ vertex may also arise from higher scale physics.

Armed with the above  we obtain finite contributions to the $\ell,\ell^\prime$ elements of the active neutrino mass matrix $M^\nu$. These can be written as
\begin{equation}
\label{eq:numass}
(M^\nu)_{\ell \ell^\prime}=\sum_{i,j}K^{ij}\lambda_\ell^{i} \lambda_{\ell^\prime}^{j},
\end{equation}
where $i,j=u,c,t$. The $K^{ij}$ factor controls the scale of neutrino masses and it is given by
\small
\begin{eqnarray}
\label{eq:Kfac}
 K^{ij}&=&\frac{y_\psi^{i} y^{j}_\psi \rho}{(16\pi^2)^3} \frac{m_i m_j\, M^{3}_{N_{R}}}{(m_\chi^2-m_i^2)(m^2_{\chi}-m_j^{2})}I(m_\psi ^2,m_\chi^{2},m_i^{2},m_j^{2}),\nonumber \\
 I(m_\psi^{2},m_\chi^{2},m_i^{2},m_j^{2})&=& \int_0^{\infty} du \, \frac{u}{u+1}f(u,m_{i}^2,m_{\psi}^2,m_{\chi}^2)f(u,m_{j}^2,m_{\psi}^2,m_{\chi}^2) \nonumber \\
 f(u,m^2,m_\psi^{2},m_\chi^{2})&=&\int_0^{1}dx\ln\left(\frac{m_\chi^{2}(1-x)+ m_\psi^{2}x +M^{2}_Nu x(1-x)}{m ^{2}(1-x)+ m_\psi^{2}x +M^2_{N}u x(1-x)}\right).
 \end{eqnarray}
 \normalsize
This reveals the workings of a generalized seesaw mechanism. The seesaw scale here is $m_\chi$. In the limit $M_{N_R}\rightarrow 0$ there is a conserved lepton number and therefore the active neutrinos will remain massless. Dimensional arguments give the other mass ratios in Equation(\ref{eq:Kfac}) as the integrals are dimensionless.
\begin{figure}[ht]\centering
\subfigure{\includegraphics[width=3.0in]{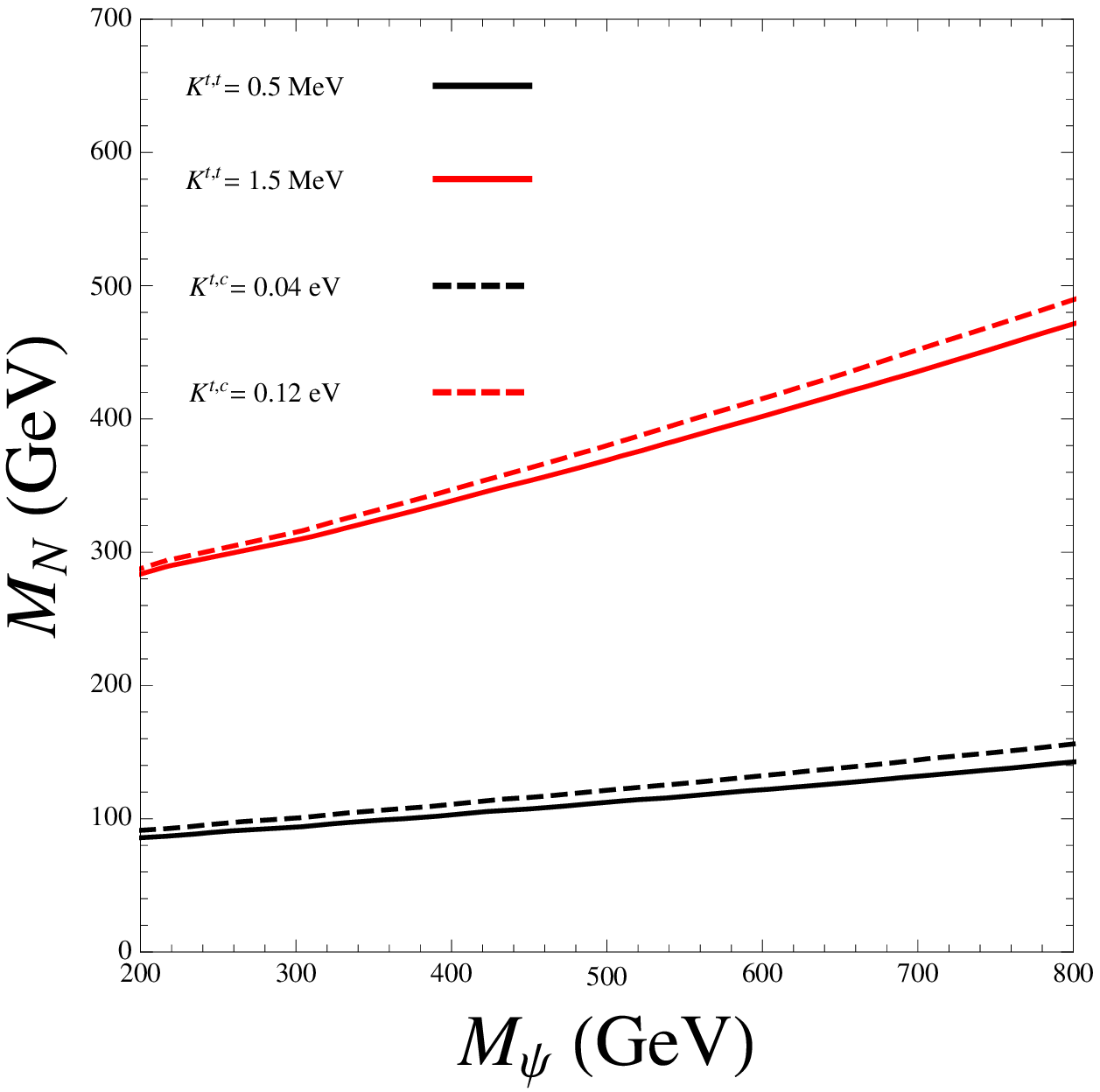}\label{fig:NeutMass_G1}}~
\subfigure{\includegraphics[width=3.0in]{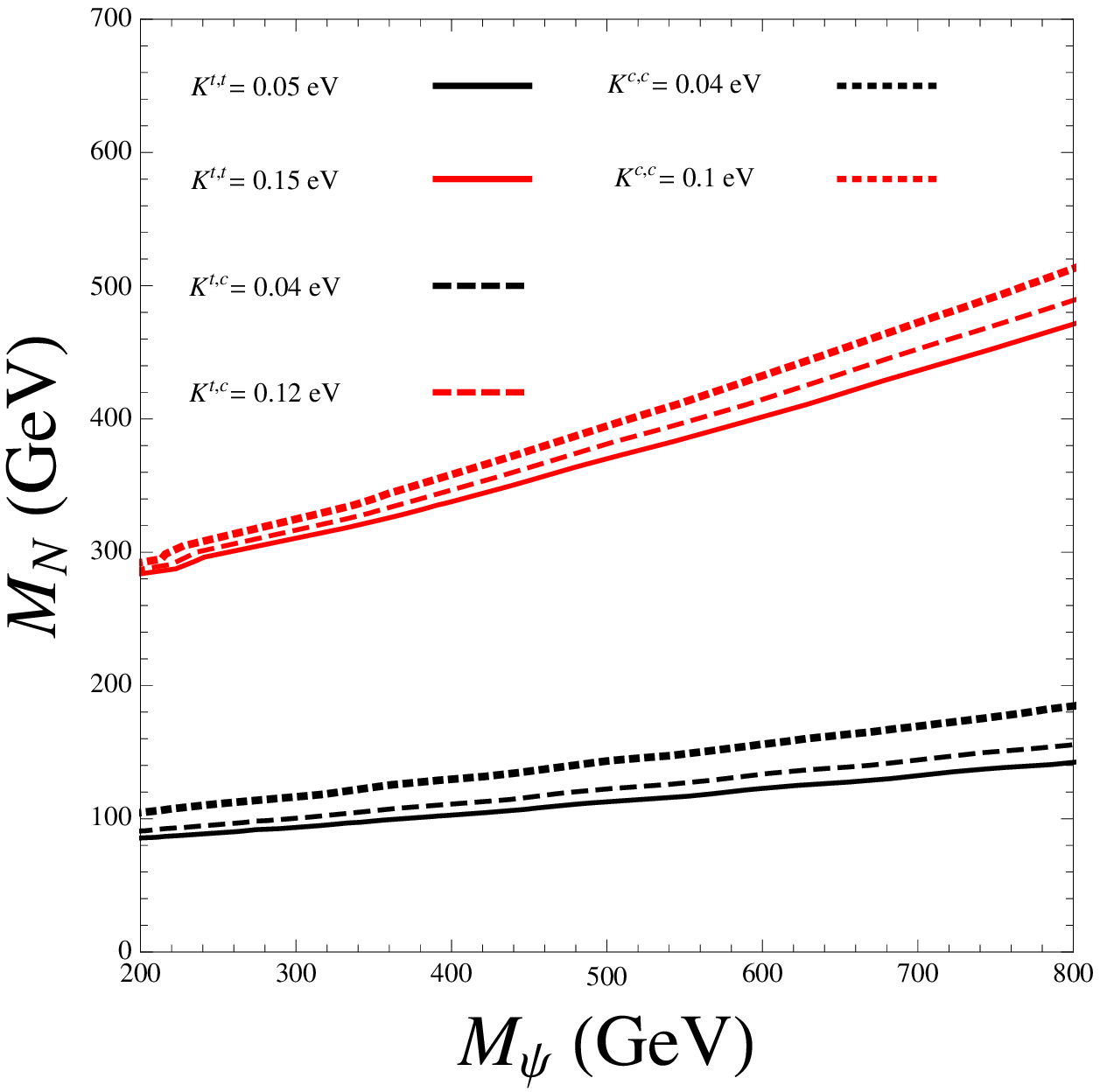}\label{fig:NeutMass_G2}}
\caption{\small $K^{t,t}$, $K^{t,c}$ and $K^{c,c}$ factors in the $M_{N_{R}}-m_{\psi}$ plane. (a)Left: $y^{t}_{\psi}=1$ and $y^{c}_{\psi}=0.01$. The red and black solid lines correspond to $K^{t,t}=1.5,0.5$ MeV respectively while the red and black dashed lines correspond to $K^{t,c}=0.12,0.04$ eV respectively. (b)Right: $y^{t}_{\psi}=0.01$ and $y^{c}_{\psi}=1$. The red and black solid lines correspond to $K^{t,t}=0.15,0.05$ eV respectively while the red and black dashed lines correspond to $K^{t,c}=0.12,0.04$ eV respectively. On the right we see that $K^{c,c}$ is of the same order as $K^{t,t}$ and $K^{t,c}$, in black and red dotted lines. }
\end{figure}

Numerically it is easy to see that the important contributions come from the $t$- and $c$-quarks without fine tuning of the Yukawa couplings. Moreover, if only one generation of quarks contributes, such as the $t$-quark, then it will give rise to two massless active neutrinos which is ruled out by neutrino oscillation data. Thus, at least two quark generations must come into play. This will result in one massless neutrino, which is consistent with current data. If all three neutrinos were to be found to have non zero masses, then the $u$-quark must also be included. We note that another solution for light neutrino masses would be to add one or two more $N_R$ but we don't pursue this alternative here.

In this work we update the results presented in~\cite{ModelMain} in various interesting regions of parameter space. We choose to vary only the mass of the coloured electroweak-singlet scalar and the Majorana neutrino, $N_{R}$. We use a coloured electroweak-triplet scalar with mass $m_{\chi}=1$ TeV and an effective coupling $\rho=0.1$ as benchmark points. In Figure~\ref{fig:NeutMass_G1} we present the $K^{t,t}$ and $ K^{t,c}$ factors in the $M_{N_{R}}-m_{\psi}$ plane for $y^{t}_{\psi}=1$ and $y^{c}_{\psi}=0.01$. The red and black solid lines correspond to $K^{t,t}=1.5,0.5$ MeV respectively. For this coupling values, $K^{t,c}$ is negligible compared to $K^{t,t}$ and this is depicted by the red and black dashed lines which correspond to $K^{t,c}=0.12,0.04$ eV respectively. However for larger values of $y^{c}_{\psi}$ and suppressed values of $y^{t}_{\psi}$, $K^{t,c}$ and $K^{c,c}$ become dominant contributions to the neutrino mass matrix and are of the same order as $K^{t,t}$. An example is depicted in Figure~\ref{fig:NeutMass_G2} which corresponds to $y^{t}_{\psi}=0.01$ and $y^{c}_{\psi}=1$.

\section{Constraints}\label{sec:constraints}
In the following subsections we discuss the main constraints on our model. In particular, we look at the regions of parameter space excluded by lepton flavour violating decays, rare decays of the top quark and the existence of new modes for Higgs decay and production. Furthermore, we look at the latest collider searches for dark matter by the CMS collaboration in the jets + MET and monojet channels.
\subsection{$\mu\to e \gamma$ and rare $b$ decays}\label{subs:mutoeg}

 Although the dark matter calculation is not sensitive to the masses of the coloured electroweak-triplet states, they can give rise to lepton flavour violating decays such as $\mu \to e \gamma$ as well as a contribution to the muon anomalous magnetic moment, $a_\mu$. Both contributions come in at the 1-loop level. Interestingly, the singlet state $\psi$ does not contribute to these processes at this level. The Feynman diagrams for the $\mu \to e \gamma $ decay process are  depicted in Figure {\ref{fig:mueg}.
\begin{figure}[ht]\centering
\includegraphics[width=5.0in]{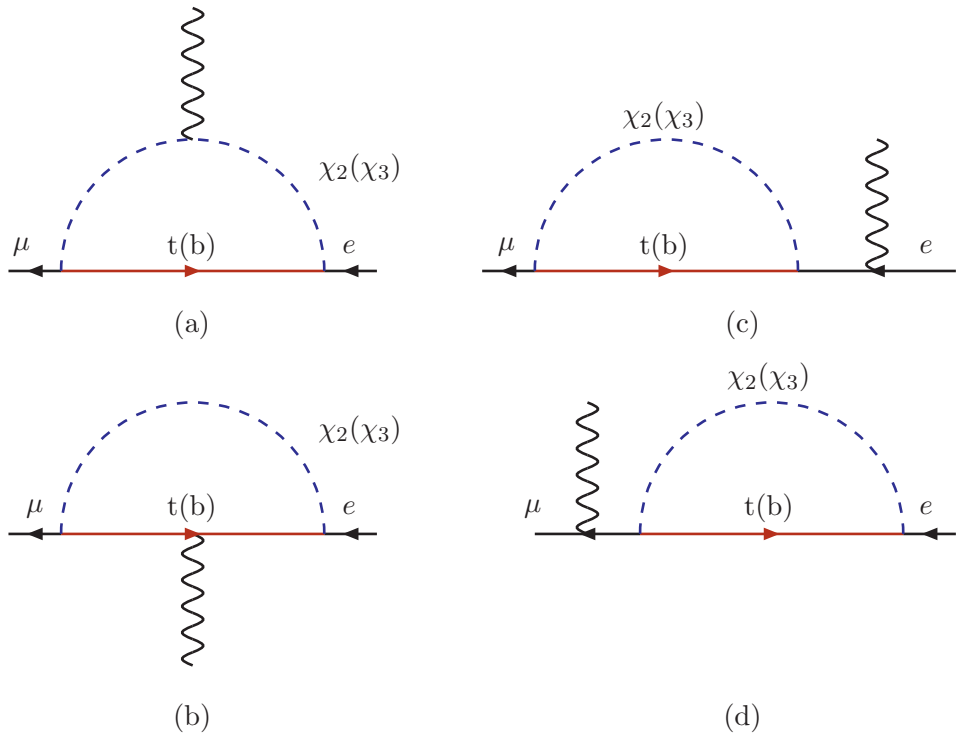}
\caption{\small 1-loop diagrams for $\mu(p) \to e(p^\prime) + \gamma(q)$ decays. The arrows indicate fermion charge flow. Main contribution comes from (a) and (b) whereas (c) and (d) are needed to enforce gauge invariance. Similar diagrams from the second generation quarks are not displayed}\label{fig:mueg}
\end{figure}

The effective Lagrangian for the decay can be written as
\begin{equation}
\mathcal{L} = A \bar{e} i\sigma^{\mu\nu}(1+\gamma^{5})\mu F_{\mu\nu}.
\end{equation}
 The decay width is given by
\begin{equation}
\label{eq:muegwidth}
\Gamma (\mu\to e \gamma) = \frac{|A|^2 m_\mu ^3}{16\pi},
\end{equation}
where a standard calculation yields the following expression for $A$
\begin{equation}
\begin{split}
\label{eq:A}
A&=\frac{e m_\mu}{64\pi^2 m_\chi ^2}\sum_{i=t,c,u}\frac{\lambda_\mu ^i \lambda_e ^i}{(1-a_i)^3}\left[ 3-10 a_i -5 a_i^2 +\frac{2a_i -16a_i ^2}{1-a_i}\ln a_i\right] \\
&\underset{a_i\to 0}{\longrightarrow} \frac{3e}{64 \pi^2}\left( \frac{m_\mu}{m_\chi ^2}\right)\sum_{i=t,c,u}\lambda_\mu ^i \lambda_e ^i .
\end{split}
\end{equation}
In the above equation we have assumed a common mass, $m_\chi$, for both $\chi_2$ and $\chi_3$ and defined $a_i=\frac{m_i^2}{M_\chi ^2}$. We have also neglected terms of $O\left((m_{t(b)}/M)^2\right)$. Therefore, the branching fraction is given by
\begin{equation}
Br(\mu\to e \gamma) =1.8\left( \frac{ \mathrm{TeV}}{m_\chi}\right)^4 \times 10^{-6}|\lambda_\mu ^t\lambda_e ^t +\lambda_\mu ^c\lambda_e ^c+\lambda_\mu ^u\lambda_e ^u|^2.\label{eq:Brmutoeg1}
\end{equation}

In the region where $K^{t,t}\gg K^{c,t},K^{c,c}$, our calculations are not sensitive to the value of $\lambda^{c}_{\mu}\lambda^{t}_{e}$ and $\lambda^{c}_{\mu}\lambda^{c}_{e}$ which appear in Equation~(\ref{eq:numass}). Furthermore, we can set $\lambda^{u}_{l}\sim0$ in order to suppress rare kaon decays. In order to analyze the constraints arising from this rare decay, we maximize the contribution from new physics by working in the limit where $\lambda^{c}_{\mu}\lambda^{c}_{e}\sim\lambda^{t}_{\mu}\lambda^{t}_{e}$ and set $m_{\chi}=1$ TeV. In order to extract an upper bound on the value of $\lambda^{t}_{\mu}\lambda^{t}_{e}$ we follow the analysis in~\cite{ModelMain} where we made use of the latest best fit value for $M^{\nu}_{e\mu}$~\cite{Bertuzzo:2013ew} assuming a normal hierarchy with $m_{1}\to0$, and the current experimental bound on $Br\left(\mu\to e\gamma\right)\le2.4\times10^{-12}$~\cite{Adam:2011ch}. With this in mind, in the limit where $y^{t}_{\psi}\gg y^{c}_{\psi}$ we can rewrite Equation~(\ref{eq:Brmutoeg1}) using Equation~(\ref{eq:numass}):
\begin{equation}
Br\left(\mu\to e\gamma\right)=7.2\times 10^{-6}\left(\frac{M^{\nu}_{e\mu}}{K^{t,t}}\right)^{2}.
\label{eq:Brmutoeg2}
\end{equation}
This limit is justified since in order to suppress contributions to the $D^{0}-\bar{D^{0}}$ oscillation parameters one must work in the limit where either $y^{c,u}_{\psi}$ are small. This is covered in Section~\ref{subs:oscillations}.

Similar diagrams to those contributing to $\mu\to e\gamma$ contribute to the decay $b\to s\gamma$. This decay has a SM contribution which is the same as the amplitude arising from our new physics. The data~\cite{Amhis:2012bh} is consistent with the SM expectation and we obtain the following limit:
\begin{equation}
\sum_{l}|\lambda^{b}_{l}\lambda^{s}_{l}|<0.104\left(\frac{m_{\chi}}{\text{TeV}}\right),
\end{equation}
where we have set $\lambda^{b,s}_{l}=\lambda^{t,c}_{l}$. Furthermore, recent results from the LHCb and CMS collaborations  have found a branching ratio for the decay $B^{0}_{s}\to\mu^{+}\mu^{-}$ of $2.9^{+1.1}_{-1.0}\times10^{-9}$~\cite{LHCB} and $3.0^{+1.0}_{-0.9}\times10^{-9}$~\cite{CMSbmm} consistent with the SM expectation of $3.2\pm0.2\times10^{-9}$. Within our framework, this decay can be induced through the exchange of the coloured electroweak-triplet, $\chi_{3}$, and it sets the following bound on the product $\lambda^{b}_{\mu}\lambda^{s}_{\mu}$:
\begin{equation}
\lambda^{b}_{\mu}\lambda^{s}_{\mu}<3.4\times10^{-3}\left(\frac{m_{\chi}}{\text{TeV}}\right).
\end{equation}
In~\cite{ModelMain} we found that values of $\lambda^{i}_{l}$ below $0.1$, required for sub-eV neutrino masses, are consistent with constraints from rare $b$ decays.

\subsection{Rare top decays}\label{subs:raretop}

Within our framework both types of Yukawa couplings, $\lambda^{i}_{l}$'s and $y^{i}_{\psi}$'s, are involved in the mechanism that leads to neutrino masses. In Section~\ref{subs:mutoeg} we saw that the decay $\mu\to e\gamma$ can directly constrain $\lambda^{i}_{\mu}\lambda^{i}_{e}$, in particular $\lambda^{t}_{\mu}\lambda^{t}_{e}$ which is inversely proportional to $(y^{t}_{\psi})^{2}$. However, the following rare decays of the top, $t\to g~c$, $t\to \gamma~c$ and $t\to Z~c$ will be sensitive to $y^{c}_{\psi}$. The dominant decay mode is $t\to g~c$ since this has a large colour charge. This mode can be probed at the LHC through single top production via gluon and $c$-quark fusion. The SM contributions are negligible and this mode can provide a very sensitive probe for new physics. In particular, the gluon plus $c$ mode is useful for probing new colour degrees of freedom such as the $\psi$ and $\chi$ scalars. In our framework, due to the special role that $\psi$ plays in the annihilation of the dark matter candidate, $N_{R}$, we consider coloured electroweak-singlet states that are much lighter than the coloured electroweak-triplet scalars. The effective Lagrangian for this decay is given by
\begin{equation}
\mathcal{L} = A_g^a \bar{c} i\sigma^{\mu\nu}P_R t G_{\mu\nu}^a,
\end{equation}
where $G_{\mu\nu}$ is the gluon field tensor. The dipole form factor $A_g$ can be calculated from diagrams similar to those that contribute to $\mu\to e\gamma$ and it is given by:
\begin{equation}
A_g^{a}=i\frac{y^t_{\psi}y^{*c}_{\psi}g_s}{16\pi^2}\frac{m_t}{m_\psi^{2}} T^a I(x_N,x_t),
\end{equation}
where $g_s$ is the QCD coupling, $T^a$ is a colour $SU(3)$ generator and $x_i= \frac{m_i^2}{M_\psi^{2}}$ for a particle of mass $m_i$. The integral $I$ is given explicitly by
\begin{equation}
I(a,b)=\int_0^{1} dx\, \int_0^{1-x} dy \; \frac{1-x-y}{x+y +a(1-x-y)-bx(1-x-y)},
\end{equation}
where the $c$-quark mass can be neglected at these energies. For the case of relatively light Majorana neutrinos, i.e. $x_{N_{R}}<<1$, the integral can
be expressed in analytic form
\begin{eqnarray}
I(0,b)=\frac{1}{6b}\left[ -6\ln(1-b)(1-b+b\ln b) +b(-6 +\pi^2 -3b +3\ln^2 b) +6b {\mathrm {Li}}_2 (1-b^{-1})\right]. \nonumber \\
\end{eqnarray}

A recent ATLAS analysis searching for flavour changing neutral currents in single top quark production with an integrated luminosity of $14.2$ fb$^{-1}$ at a centre of mass energy of $\sqrt{s}=8$ TeV~\cite{TheATLAScollaboration:2013vha} can be used to place an upper bound on the parameters of our model. This bound is given by
\begin{equation}
\frac{y_\psi^{t}y_\psi^{c}m_t}{32\pi^2 m_\psi^2} I(x_N,x_t)< 1.1 \times 10^{-2} \mathrm {TeV}^{-1}.
\label{eq:tcg}
\end{equation}

\subsection{$D^{0}-\bar{D}^{0}$ Oscillations}\label{subs:oscillations}

Within our framework, both $u$- and $c$-quarks can couple to the coloured electroweak-singlet and the Majorana neutrino for non-zero values of $y^{u,c}_{\psi}$. Couplings of this type can yield sizeable contributions to the mass difference, $\Delta M_{D}$, in $D^{0}-\bar{D}^{0}$ mixing.
Meson-antimeson mixing is sensitive to heavy degrees of freedom and as such can significantly constrain the validity of our model. The relevant quantities in $D^{0}-\bar{D}^{0}$ mixing are the mass difference, $\Delta M_{D}$, and the width difference, $\Delta \Gamma_{D}$, that can be parametrized by the following equations:
\begin{eqnarray}
x_{D}&=&\frac{\Delta M_{D}}{\Gamma_{D}} \nonumber \\
y_{D}&=&\frac{\Delta \Gamma_{D}}{2\Gamma_{D}},
\end{eqnarray}
where $\Gamma_{D}$ is the average width of the two neutral $D$ meson mass eigenstates. A current fit to these two variables by the HFAG collaboration gives $x_{D}=0.43^{+0.15}_{-0.16}\%$, $y_{D}=0.65\pm0.08\%$~\cite{Amhis:2012bh}.

Given the uncertainty in the SM long distance contribution to $\Delta M_{D}$, we assume that $\Delta M_{D}$ is driven primarily by contributions from this model. This gives us a tighter constraint on our model than assuming that the SM long distance contribution is comparable to the size of $\Delta M_{D}$. Following the analysis of the implications of $D^{0}-\bar{D}^{0}$ mixing for new physics~\cite{Golowich:2007ka}, our contribution to $x_{D}$ is given by
\begin{equation}
x_{D}=\frac{1}{M_{D}\Gamma_{D}}Re\left[2\left<\bar{D}^{0}|H^{|\Delta C|=2}_{NP}|D^{0}\right>\right],
\end{equation}
where we can use the operator product expansion and the renormalization group to define
\begin{equation}
\left<\bar{D}^{0}|H^{|\Delta C|=2}_{NP}|D^{0}\right>=G\sum_{i=1}C_{i}(\mu)\left<\bar{D}^{0}|Q_{i}|D^{0}\right>(\mu).
\end{equation}
In the above equation $G$ is a coefficient with inverse squared mass dimension, $C_{i}$ are Wilson coefficients, and $\left<\bar{D}^{0}|Q_{i}|D^{0}\right>$ are effective operators. In our framework, the Lagrangian in Equation~\ref{eq:lag} yields the following operator at the scale where the heavy degrees of freedom are integrated out,
\begin{equation}
Q_{6}=\left(\bar{u}_{R}\gamma_{\mu}c_{R}\right)\left(\bar{u}_{R}\gamma^{\mu}c_{R}\right).
\end{equation}
To take into account the operator mixing between $Q_{6}$ and the other $7$ operators listed in~\cite{Golowich:2007ka} that arise from the renormalization group running between the scale of new physics and the charm mass $m_{c}$, we solve the renormalization group equations obeyed by the Wilson coefficients. With this in mind, we obtain a contribution to the mass difference given by
\begin{equation}
\Delta M_{D}=\frac{\left(y^{u}_{\psi}y^{c}_{\psi}\right)^{2}f_{D}M_{D}}{64\pi^{2}m_{\psi}}\frac{2}{3}B_{D}\beta\left(m_{c},m_{\psi}\right)|\L(\eta)|,\label{eq:ddBound}
\end{equation}
where we have used the values $f_{D}=212\times 10^{-3}$ GeV and $B_{D}=0.82$ for the $D$-meson decay constant and Bag constant respectively. Furthermore, we have used an average $D$-meson mass of $M_{D}=1.865$ GeV, a renormalization group factor $\beta\left(m_{c},m_{\psi}\right)$ given by
\begin{equation}
\beta\left(m_{c},m_{\psi}\right)=\left(\frac{\alpha_{s}(m_{\psi})}{\alpha_{s}(m_{t})}\right)^{2/7}\left(\frac{\alpha_{s}(m_{t})}{\alpha_{s}(m_{b})}\right)^{6/23}\left(\frac{\alpha_{s}(m_{b})}{\alpha_{s}(m_{c})}\right)^{6/25},
\end{equation}
and the loop integral factor $\L(\eta)$ given by
\begin{equation}
\L(\eta)=\frac{\eta}{(1-\eta)^{2}}\left[1+\frac{1}{(1-\eta)}\log\eta\right],
\end{equation}
with $\eta=\frac{M^{2}_{N_{R}}}{m^{2}_{\psi}}$. With this is mind we exclude regions of parameter space consistent with
\begin{equation}
\frac{\Delta M_{D}}{\Gamma_{D}}>0.43\%,
\end{equation}
where $\Gamma_{D}$ is the width of the neutral $D$ meson given by $\Gamma_{D}=1.605\times 10^{-12}$ GeV.
\subsection{Higgs production and decay}\label{sec:higgspd}

The discovery of a $125$ GeV SM-like Higgs boson at the LHC~\cite{Aad:2012tfa,Chatrchyan:2012ufa} has placed strong bounds on models that modify how the Higgs is produced and how it decays. In our model, the new coloured scalar degrees of freedom contribute, at one loop, to SM Higgs production through gluon fusion, and Higgs decays into photons. Since in our framework the coloured electroweak-singlet, $\psi$, may lie below the TeV scale, the relevant operator contributing to Higgs production and decay after electroweak symmetry breaking (EWSB) is given by
\begin{equation}
\kappa_{3}H^{\dagger}H\psi^{\dagger}\psi\to \kappa v H\psi^{\dagger}\psi,
\end{equation}
where $v=\left<H\right>=174$ GeV. The LHC is now able to restrict the couplings of these new coloured scalars to the SM-like Higgs boson. An analysis in~\cite{Chang:2012ta} studied the contributions to the Higgs decay width into photons that arise from coloured scalars. Using standard notations, the Higgs diphoton decay width, including only new spin-0 contributions, is given by
\begin{equation}
\Gamma_{\gamma\gamma}\equiv\Gamma\left(H\to\gamma\gamma\right)=\frac{G_{\mu}\alpha^{2}M^{3}_{H}}{128\sqrt{2}\pi^{3}}\left| F_{1}(\tau_{W})+\frac{4}{3}F_{1/2}(\tau_{t})+d(r_{\psi})Q^{2}_{\psi}\frac{\kappa_{3}}{g_{w}}\frac{M^{2}_{W}}{m^{2}_{\psi}}F_{0}(\tau_{\psi})\right|^{2},
\end{equation}
where $Q_{\psi}=2/3$ and $d(r_{\psi})=3$ are the charge and dimension of the representation of the coloured electroweak-singlet. The functions $F_{1},~F_{1/2}$ and $F_{0}$ are given by
\begin{eqnarray}
F_{0}(\tau)&=&-[\tau-f(\tau)]\tau^{-2}\nonumber \\
F_{1/2}(\tau)&=&2[\tau+(\tau-1)f(\tau)]\tau^{-2} \nonumber \\
F_{1}(\tau)&=&-[2\tau^{2}+3\tau+3(2\tau-1)f(\tau)]\tau^{-2},
\end{eqnarray}
where
\begin{equation}
f(\tau)=\begin{pmatrix}
\arcsin^{2}\sqrt{\tau}~~~~~~~~~~~~~~~~~~~~~\tau\le1 \\
-\frac{1}{4}\left[\log\frac{1+\sqrt{1-\tau^{-1}}}{1-\sqrt{1-\tau^{-1}}}-i\pi\right]^{2}~~~\tau>1
\end{pmatrix}
\end{equation}
and $\tau_{i}=M^{2}_{H}/4M^{2}_{i}$.

The analysis on additional contributions to the production of a SM-like Higgs boson through gluon fusion was carried out in a similar fashion to the diphoton Higgs decay~\cite{Chang:2012ta}. In particular, the parton level cross section for $gg\to H$ is given by
\begin{equation}
\sigma_{gg}\equiv\hat{\sigma}(gg\to H)=\sigma_{0}M^{2}_{H}\delta(\hat{s}-M^{2}_{H}),
\end{equation}
where
\begin{equation}
\sigma_{0}=\frac{G_{\mu}\alpha^{2}_{s}}{128\sqrt{2}\pi}\left| \frac{1}{2}F_{1/2}(\tau_{t})+C(r_{\psi})\frac{\kappa_{3}}{g_{w}}\frac{M^{2}_{W}}{m^{2}_{\psi}}F_{0}(\tau_{\psi})\right|^{2}\label{eq:ggfuse},
\end{equation}
and $C(r_{\psi})=1/2$ is the index of the representation of $\psi$.

\begin{figure}[ht]\centering
\includegraphics[width=3.0in]{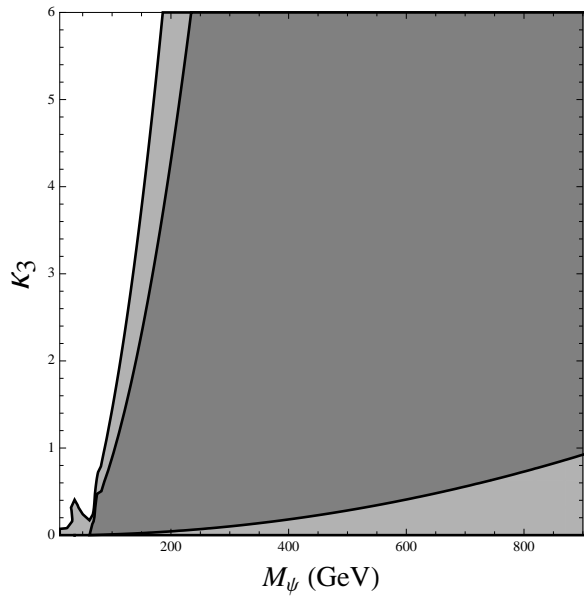}
\caption{\small Regions in the $m_{\psi}-\kappa_{3}$ plane consistent with the CMS and ATLAS Higgs data at the $1\sigma$ (light grey) and $2\sigma$ (dark grey) levels.  }\label{fig:muHiggs}
\end{figure}
Recent results from the LHC suggest a diphoton Higgs decay channel in agreement with the SM. In particular, the ATLAS collaboration measured a $\sigma/\sigma_{SM}=1.32\pm0.38$~\cite{Aad:2014eha}. The CMS result of $1.14^{+0.26}_{-0.23}$~\cite{Khachatryan:2014ira} . We use these experimental results on the signal strength in the gluon fusion production mode and decay to two photons together with the mass of the Higgs measured by ATLAS and CMS of $125.40\pm0.37~\text{(stat)}\pm0.18~\text{(syst)}$ and $124.70\pm0.34$ GeV respectively and find $2\sigma$ and $1\sigma$ contours on the $m_{\psi}-\kappa_{3}$ plane using a combined $\chi^{2}$ fit. Our results are shown in Figure~\ref{fig:muHiggs}.

\subsection{Collider Constraints}\label{sec:collider}

One important feature of our model is that it contains new coloured degrees of freedom which can be produced in hadron colliders such as the LHC. In particular, the coloured electroweak-singlet, $\psi$, can be pair produced and it can later decay to top and/or light up-type quarks and a large component of missing transverse energy carried away by $N_{R}$. Additionally, it can be singly produced in association with $N_{R}$ leading to a monojet or monotop signal. A monojet signal is also viable through pair production of Majorana neutrinos with a jet emitted from an intermediate coloured electroweak-singlet, $\psi$. In what follows we discuss the different searches carried out by the CMS collaboration used to constrain the parameter space considered in this work. The simulation of the signal at the parton level is carried out using MadGraph 5~\cite{Alwall:2011uj}} with model files generated with FeynRules~\cite{Feynr}. The parton showering and hadronization are carried out with PYTHIA~\cite{Sjostrand:2006za} and the detector simulation using Delphes 3~\cite{delphes}. The Delphes parameters are changed according to the collider analyses and are discussed below. In addition, we carry out a channel by channel exclusion using a $95\%$ CL excluded number of signal events calculated using a single-channel $CL_{s}$ method adapted from the CheckMATE program~\cite{Drees:2013wra}. We choose three benchmark scenarios defined by three sets of fixed couplings $(y^{t}_{\psi},y^{c}_{\psi},y^{u}_{\psi})=(1,0.1,0.1),(1,0.01,0.5),(0.4,0.01,1)$. These scenarios are motivated by the constraints discussed in the previous subsections, in particular, a suppression of $D^{0}-\bar{D}^{0}$ oscillations, a small contribution to the decay of $\mu\to e\gamma$ and natural neutrino masses. We then compare our excluded regions in the $M_{N_{R}}-m_{\psi}$ plane with the exclusions generated with CheckMATE which combines the following validated ATLAS and CMS analyses:
\begin{itemize}
\item
$1~\text{lepton}+4~\text{jets}+\slashed{E}_{T}$~\cite{ATLAS:2012tna}
\item
$\text{Monojet search}+\slashed{E}_{T}$~\cite{ATLAS:2012zim}
\item
$0~\text{lepton}+6~(2b)\text{-jets}+\slashed{E}_{T}$\cite{ATLAS:2013cma}
\item
$2-6~\text{jets}+\slashed{E}_{T}$\cite{TheATLAScollaboration:2013fha}
\item
$2~\text{leptons}+~\text{jets}+\slashed{E}_{T}~(razor)$\cite{TheATLAScollaboration:2013via}
\item
$\text{At least 2 jets}+b~\text{jet multiplicity}+\slashed{E}_{T}(\alpha_{T})$\cite{Chatrchyan:2013lya}

\end{itemize}

\subsubsection{Limits from jets+MET}\label{subsub:jets}

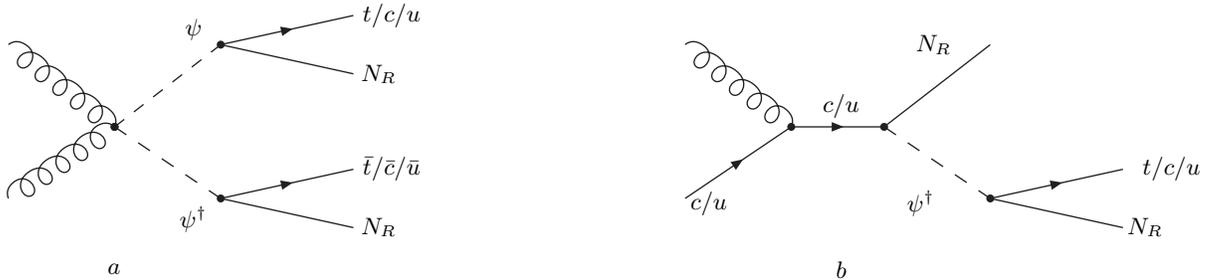
\begin{figure}[!ht]
\begin{center}\begin{picture}(410,100)(0,-60)
\small
\Gluon(0,10)(40,-21){4}{5}
\Vertex(40,-21){1.5}
\Gluon(0,-48)(40,-21){4}{5}
\Vertex(40,-21){1.5}
\DashLine(80,10)(40,-21){5} \Text (70,16)[]{$\psi$}
\DashLine(80,-48)(40,-21){5} \Text (70,-56)[]{$\psi^{\dagger}$}

\Vertex(80,10){1.5}
\ArrowLine(80,10)(130,21) \Text (145,21)[]{$t/c/u$}
\Line(80,10)(130,-1) \Text (140,-1)[]{$N_{R}$}

\Vertex(80,-48){1.5}
\ArrowLine(80,-48)(130,-37) \Text (145,-37)[]{$\bar{t}/\bar{c}/\bar{u}$}
\Line(80,-48)(130,-59) \Text (140,-59)[]{$N_{R}$}
\Text(40,-75)[]{{$a$}}


\Gluon(255,10)(295,-21){4}{5}
\Vertex(295,-21){1.5}
\ArrowLine(255,-48)(295,-21) \Text (265,-50)[]{$c/u$}

\ArrowLine(295,-21)(330,-21) \Text (315,-13)[]{$c/u$}

\Vertex(330,-21){1.5}
\Line(330,-21)(370,10) \Text (350,10)[]{$N_{R}$}
\DashLine(330,-21)(370,-48){5} \Text (345,-50)[]{$\psi^{\dagger}$}

\Vertex(370,-48){1.5}
\ArrowLine(370,-48)(420,-37) \Text (440,-37)[]{$t/c/u$}
\Line(370,-48)(420,-59) \Text (430,-59)[]{$N_{R}$}
\Text(315,-75)[]{{$b$}}

\end{picture}
\end{center}
\caption{\small Leading order Feynman diagrams the lead to jets plus MET final states at the LHC. The diagram on the left will also lead to a $t\bar{t}$ plus MET signature and can be constrained by searches for scalar top pair production~\cite{Chatrchyan:2013xna}.}
\label{fig:multijet1}
\end{figure}
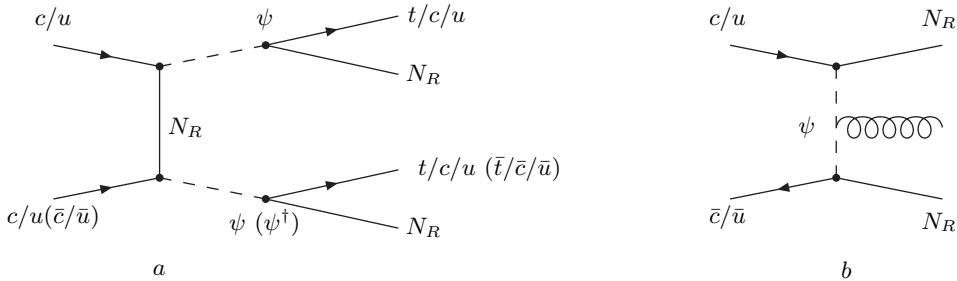
\begin{figure}[!ht]
\begin{center}\begin{picture}(410,100)(0,-60)
\small
\ArrowLine(0,10)(40,2)\Text(0,20)[]{$c/u$}
\Vertex(40,2){1.5}
\DashLine(40,2)(80,10){5} \Text (80,20)[]{$\psi$}
\ArrowLine(0,-48)(40,-40) \Text(0,-55)[]{$c/u (\bar{c}/\bar{u})$}

\Vertex(80,10){1.5}
\ArrowLine(80,10)(130,21) \Text (145,21)[]{$t/c/u$}
\Line(80,10)(130,-1) \Text (140,-1)[]{$N_{R}$}

\Line(40,-40)(40,2) \Text (50,-21)[]{$N_{R}$}

\Vertex(40,-40){1.5}
\DashLine(40,-40)(80,-48){5} \Text (80,-57)[]{$\psi$ ($\psi^{\dagger}$)}
\Vertex(80,-48){1.5}
\ArrowLine(80,-48)(130,-37) \Text (165,-37)[]{$t/c/u~(\bar{t}/\bar{c}/\bar{u})$}
\Line(80,-48)(130,-59) \Text (140,-59)[]{$N_{R}$}
\Text(40,-75)[]{{$a$}}


\ArrowLine(255,10)(295,2)\Text(255,20)[]{$c/u$}
\Vertex(295,2){1.5}
\Line(295,2)(335,10) \Text(335,20)[]{$N_{R}$}
\ArrowLine(295,-40)(255,-48) \Text(255,-55)[]{$\bar{c}/\bar{u}$}

\DashLine(295,-40)(295,2){5} \Text(285,-21)[]{$\psi$}
\Gluon(295,-21)(335,-21){4}{5}

\Vertex(295,-40){1.5}
\Line(295,-40)(335,-48) \Text(335,-57)[]{$N_{R}$}
\Text(300,-75)[]{{$b$}}

\end{picture}
\end{center}
\caption{\small  Leading order Feynman diagrams that lead to jets plus MET final states at the LHC and that enhance the production rate at large $y^{t,c,u}_{\psi}$. The diagram on the right contributes to same-sign top (anti-top) production at the LHC.}\label{fig:multijet2}
\end{figure}

In this section we look at the possibility of setting limits to our model by considering searches for jets + MET at hadron colliders. We focus on a search for multijets  and missing momentum with $19.5$ fb$^{-1}$ of data at $\sqrt{s}=8$ TeV by the CMS collaboration~\cite{CMS:2013gea}. In our framework there are various topologies that lead to a multijet plus missing momentum final state, these are depicted in Figures~\ref{fig:multijet1} and~\ref{fig:multijet2}. For small couplings, $y^{c,u}_{\psi}$, the dominant channel is depicted in~\ref{fig:multijet1}($a$). In fact, this diagram resembles pair production of scalar quarks (squarks) in supersymmetry with a final state containing between two and six jets for arbitrary choices of the couplings $y^{t,c,u}_{\psi}$. For large $y^{c,u}_{\psi}$ couplings, the production cross section is enhanced via the diagram depicted in Figure~\ref{fig:multijet2}($a$) since a Majorana fermion mediates the reaction. This enhancement is more pronounced for large $M_{N_{R}}$.

In the analysis, all reconstructed particles are clustered into jets using the anti-$k_{T}$ clustering algorithm with a cone size $\Delta R=0.5$. The effects of pile up on the missing energy, in addition to those implemented in Delphes, are taken into account using a gaussian function to smear the amount of missing energy based on the total $p_{T}$ in the event~\cite{AtlasSmear}. Events containing isolated electrons or muons with $p_{T}>10$ GeV are vetoed. The selection criteria is on the number of jets, $N_{jets}$, with a transverse momentum of $p_{T}>50$ GeV and a pseudorapidity of $|\eta|<2.5$, the visible hadronic activity, $H_{T}=\sum_{N_{jets}}|p_{T}|$ for jets with $p_{T}>50$ GeV and $|\eta|<2.5$, and the momentum imbalance $\slashed{H}_{T}=|-\sum_{jets}\vec{p}_{T}|$ for jets with $p_{T}>30$ GeV and $|\eta|<5$. Furthermore, each event is required to have MET above $150$ GeV. All together, $36$ signal regions are considered.

The sensitivity of the CMS search to our model for $y^{t}_{\psi}\to1$ and small $y^{c,u}_{\psi}$ decreases dramatically since now the pair produced coloured singlet scalars will decay predominantly to $t$-quarks plus MET. In this particular case, the signal region most sensitive to our model is one where $N_{jets}=6-7$; however since the analysis does not implement a $t$-quark reconstruction algorithm, the regions with the highest energy jets are those where $\psi$ has a large mass and thus a smaller production cross section. The signal regions where $N_{jets}=3-5$ lead to a large number of signal events but not enough to overcome the large systematic uncertainties in the calculated background, mainly the contribution from QCD. The sensitivity of the CMS search to our model in the large $y^{c,u}_{\psi}$ limit is strongest for large Majorana masses and away from the degenerate region $m_{\psi}\approx M_{N_{R}}$, where high $p_{T}$ jets from the decay of the coloured electroweak-singlet are more prominent.

\subsubsection{Limits from monojets}

Here we discuss how we can set limits to our model by considering searches for monojets at hadron colliders. We focus on a search for monojet events with $19.5$ fb$^{-1}$ of data at $\sqrt{s}=8$ TeV by the CMS collaboration~\cite{CMS:rwa}. Within our framework the topologies that may lead to a final state with one high energy jet and missing momentum are depicted in Figures~\ref{fig:multijet1}($b$) and~\ref{fig:multijet2}($b$). For small values of the coupling $y^{c,u}_{\psi}$, the topology in Figure~\ref{fig:multijet1}($b$) dominates. This enhancement is more prominent when the coloured electroweak-singlet is produced on-shell. As the value of the $y^{c,u}_{\psi}$ coupling increases the diagram depicted in Figure~\ref{fig:multijet2}($b$) will enhance the cross section when a jet is emitted from the intermediate $\psi$ state.

In the analysis, all reconstructed particles are clustered into jets using the anti-$k_{T}$ clustering algorithm with a cone size $\Delta R=0.5$. The effects of pile up on the missing energy, in addition to those implemented in Delphes, are taken into account using a gaussian function to smear the amount of missing energy based on the total $p_{T}$ in the event~\cite{AtlasSmear}. To suppress the SM backgrounds, events with reconstructed muons with $p_{T}>10$ GeV, $|\eta|<2.1$ and electrons with $p_{T}>10$ GeV and $|\eta|<1.44$ or $1.56<|\eta|<2.5$ are rejected. In addition events with a well identified tau with $p_{T}>20$ GeV and $|\eta|<2.3$ are also vetoed. The selection criteria requires a jet with $p_{T}>110$ GeV and $|\eta|<2.4$ and a second jet with a $p_{T}>30$ GeV and $|\eta|<2.5$. The two jets require a separation of $\Delta\phi<2.5$ to suppress QCD dijet events. The analysis is then performed in seven regions of missing energy: $\slashed{E}_{T}>250,300,350,350,400,450,500,550$ GeV.

\subsubsection{Limits from top squark pair production and same-sign tops}\label{subsub:stops}

Additionally, one may place limits from searches for top squark pair production at hadron colliders. In this section we focus on a search for top squark pair production in the single lepton final state with $19.5$ fb$^{-1}$ of data at $\sqrt{s}=8$ TeV center of mass energies by the CMS collaboration~\cite{Chatrchyan:2013xna}. Within our framework, the topologies that lead to a $t\bar{t}$/$tt$ and MET final state are depicted in Figure~\ref{fig:multijet1}($a$), and Figure~\ref{fig:multijet2}($a$) for large $y^{c,u}_{\psi}$.

The sensitivity of the CMS search to our model is strongest for $y^{t}_{\psi}\to1$ and $y^{c,u}_{\psi}\to0$ since in this region of parameter space $BR(\psi\to N_{R}~t)\approx1$. It is important to note that the topologies contributing to this analysis also lead to a final state with jets plus MET. However, the former has the advantage that top tagging is implemented. This is done by demanding that three jets in the event originate from a top quark. This allows us to remove a larger amount of background without demanding jets with large $p_{T}$. In the analysis this is done by calculating a hadronic top $\chi^{2}$ variable for each triplet of jets in the event.

We implement the following selection criteria: Events are required to have one electron or muon with $p_{T}>25~(30)$ GeV and $|\eta|<1.4442~(2.1)$ and lie within a cone $\Delta R < 0.3$ centered around the lepton. In addition, events are vetoed if they contain a second lepton with $p_{T}>10$ GeV. All reconstructed particles are clustered into jets using the anti-$k_{T}$ clustering algorithm with a cone size $\Delta R=0.5$. The effects of pile up on the missing energy, in addition to those implemented in Delphes, are taken into account using a gaussian function to smear the amount of missing energy based on the total $p_{T}$ in the event~\cite{AtlasSmear}. Events are required to contain at least four jets with $p_{T}>30$ GeV and $|\eta|<2.5$. At least one jet must be consistent with a $b$-jet and we implement combined secondary vertex medium working point (CSVM) $b$-tagging algorithm~\cite{CSVM}. We require that events contain a transverse mass defined by $M_{T}=\sqrt{2\slashed{E}_{T}p^{l}_{T}(1-\cos\Delta\phi)}>120$ GeV, where $p^{l}_{T}$ is the transverse momentum of the lepton and $\Delta\phi$ is the azimuthal separation between the lepton and the missing energy. In addition, we also implement the hadronic top $\chi^{2}$ variable and apply the algorithm used in the CMS analysis. The variable is defined by
\begin{equation}
\chi^{2}=\frac{(M_{j_{1}j_{2}j_{3}}-M_{top})^{2}}{\sigma^{2}_{j_{1}j_{2}j_{3}}}+\frac{(M_{j_{1}j_{2}}-M_{W})^{2}}{\sigma^{2}_{j_{1}j_{2}}},\label{eq:chi2}
\end{equation}
where $M_{j_{1}j_{2}j_{3}}$ is the mass of the three-jet system, $M_{j_{1}j_{2}}$ is the mass of the two jet system, $M_{top}=174$ GeV, the pole mass of the top quark, and $M_{W}=80.4$ GeV the mass of the $W$ boson. The variables $\sigma_{j_{1}j_{2}j_{3}},~\sigma_{j_{1}j_{2}}$ are the uncertainties on the masses calculated from the jet energy resolution. Events are required to lie in the region where $\chi^{2}<5$. Furthermore the minimum angular separation between the MET and either of the two highest $p_{T}$ jets, $\Delta\phi^{min}_{p_{T,j},\slashed{E}_{T}}$ is required to lie below $0.8$.  In our analysis we consider four signal regions in missing transverse energy: MET$=150,200,250,300$ GeV.

In addition, the diagram depicted in Figure~\ref{fig:multijet2}($a$) yields a same-sign top final state. Same-sign top production yields a signal with two same-sign leptons which has a low background rate in the SM. Within our framework this final state is mediated by a Majorana fermion and thus is enhanced for large $M_{N_{R}}$. Since $m_{\psi}>M_{N_{R}}$ within our framework, one would equally need large $\psi$ masses with smaller production rates. Furthermore a large coupling, $y^{u}_{\psi}$, is required, but this has the effect of suppressing the decay rate $\psi\to t~N_{R}$. Therefore, we expect current searches for same-sign top quarks to have little constraining power on our parameter space. We focus on a search for new physics in events with same-sign dileptons with $19.5$ fb$^{-1}$ of data at $\sqrt{s}=8$ TeV by the CMS collaboration~\cite{SST} but we find its reach not competitive with the searches described above.

\subsubsection{Limits from monotop final states}

Recently, the CMS collaboration has performed a search for new physics in monotop final states with $19.7$ fb$^{-1}$ of data at $\sqrt{s}=8$ TeV centre of mass energies~\cite{CMS:2014hba}. Their reach extends to scalar and vectorial dark matter particles with masses below $327$ GeV and $655$ GeV respectively. These bounds are within the framework of effective field theories. Within our model, the topology leading to a monotop final state is depicted in Figure~\ref{fig:multijet1}($b$). We expect the sensitivity of this search to be enhanced for large values of $y^{u}_{\psi}$ due to a larger production cross section.

In the analysis, all reconstructed particles are clustered into jets using the anti-$k_{T}$ algorithm with a cone size of $\Delta R=0.5$. The effects of pile up on the missing energy, in addition to those implemented in Delphes, are taken into account using a gaussian function to smear the amount of missing energy based on the total $p_{T}$ in the event~\cite{AtlasSmear}. In the search three jets with $p_{T}>35$ GeV are considered and the two leading jets with $p_{T}>60$ GeV. Furthermore, the invariant mass of the three jets in the event has to be less than $250$ GeV and events with additional jets with $p_{T}>35$ GeV are vetoed. One $b$-tagged jet is required and we implement the CSVM $b$-tagging algorithm~\cite{CSVM}. Events with electrons or muons satisfying $p_{T}>20~(10)$ and $|\eta|<2.5~(2.4)$ are vetoed as well as events with MET$<350$ GeV. The analysis is performed in two signal regions defined by zero or one $b$-tagged jet.

\section{Results}\label{sec:results}

In this section we present the results from a scan on the masses for the coloured electroweak-singlet scalar and the mass of the Majorana neutrino, $N_{R}$, for three sets of fixed couplings  $(y^{t}_{\psi},y^{c}_{\psi},y^{u}_{\psi})=(1,0.1,0.1),(1,0.01,0.5),(0.4,0.01,1)$. In order to show the available parameter space consistent with the present dark matter abundance and all of the constraints discussed in the previous section, we present our results in the $m_{\psi}-M_{N_{R}}$ plane.
\begin{figure}[ht]\centering
\subfigure{\includegraphics[width=3.0in]{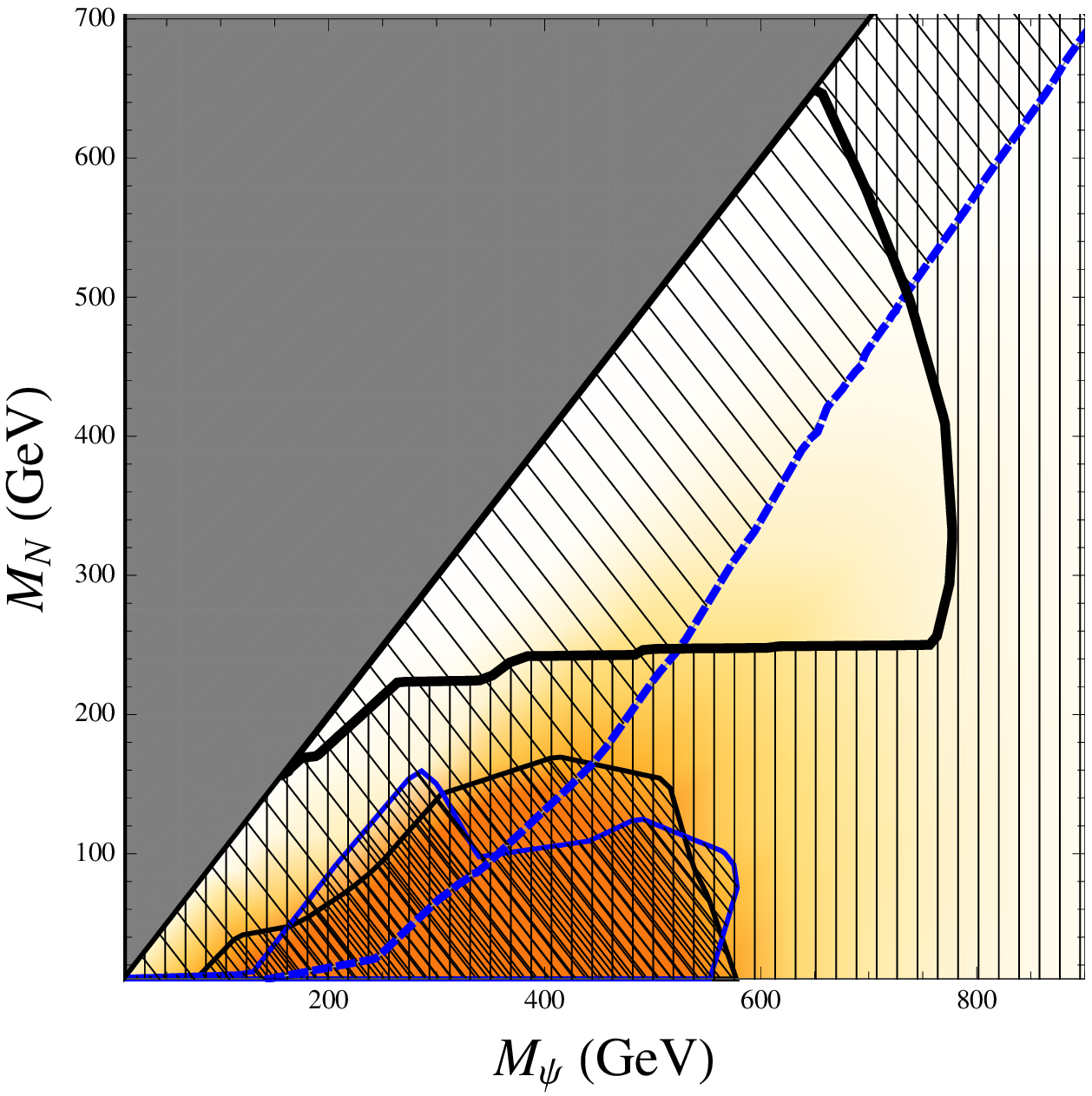}\label{fig:Results_G1}}~
\subfigure{\includegraphics[width=3.0in]{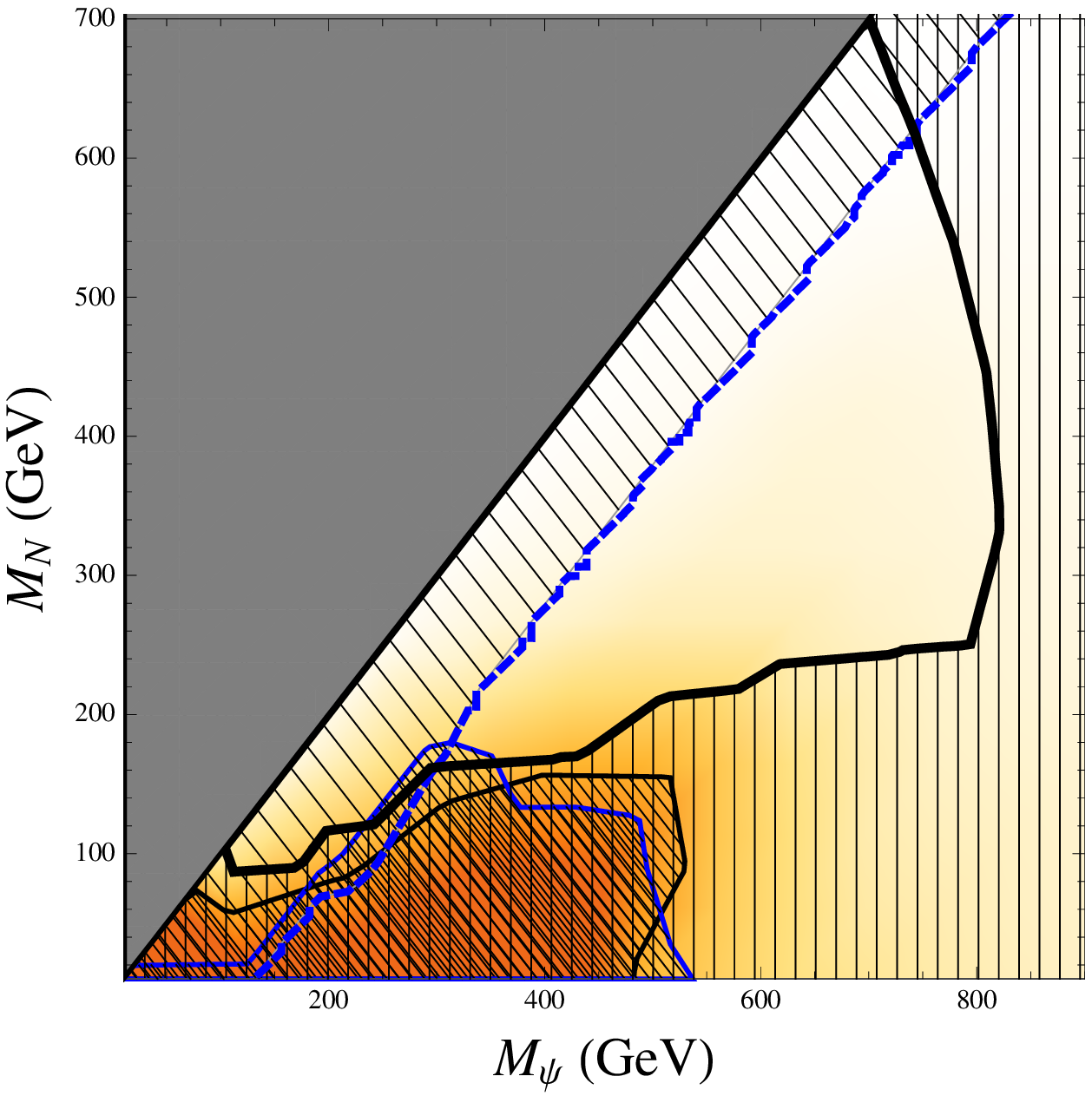}\label{fig:Results_G2}}
\caption{\small Allowed region of parameter space consistent with the density of dark matter as measured by Planck~\cite{Ade:2013zuv} (red solid line) after taking into account all constraints discussed in Section~\ref{sec:constraints} for $y^{t}_{\psi}=1$ and $y^{c,u}_{\psi}=0.1$ on the left and $y^{t}_{\psi}=1$ and $y^{c,u}_{\psi}=0.01,0.5$ on the right. The region in white is allowed but our dark matter annihilates too efficiently in the early universe.}
\end{figure}
\begin{figure}[ht]\centering
\includegraphics[width=3.0in]{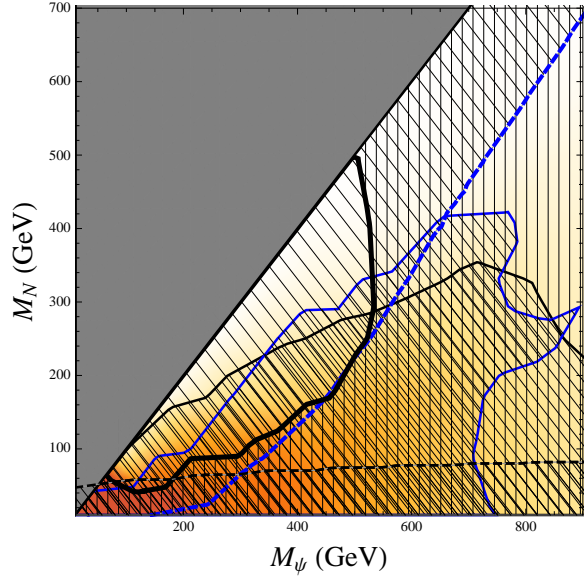}
\caption{\small Allowed region of parameter space consistent with the density of dark matter as measured by Planck~\cite{Ade:2013zuv} (red solid line) after taking into account all constraints discussed in Section~\ref{sec:constraints} for $y^{t}_{\psi}=0.4$ and $y^{c,u}_{\psi}=0.01,1$. The region in white is allowed but the dark matter annihilates too efficiently in the early universe. }\label{fig:Results_G3}
\end{figure}

In Figure~\ref{fig:Results_G1} the region consistent with the density of dark matter as measured by Planck~\cite{Ade:2013zuv} is depicted by a black thick solid line for $y^{t}_{\psi}=1$ and $y^{c,u}_{\psi}=0.1$. The dark grey region is excluded since we have assumed that $m_{\psi}>M_{N_{R}}$ in order for $N_{R}$ to be the lightest stable particle under the dark parity. The region below the thin black line is excluded from the combination of collider constraints discussed above and the thin blue line, the region excluded using CHECKMate~\cite{Drees:2013wra}. These two regions differ by at most $40$ GeV and in regions that are ruled out by a combination of overproduction of dark matter in the early universe (vertical hatched lines) and $D^{0}-\bar{D}^{0}$ oscillations, above the blue dashed line. The unhatched region is consistent with all experimental constraints, but fails to yield $100\%$ of the dark matter relic abundance. The coloration is the sensitivity on the uncertainty on the number of signal events and can also suggest the sensitivity to an increase in luminosity. Similarly, in Figure~\ref{fig:Results_G2} the region consistent with the density of dark matter is depicted by a thick solid black line for $y^{t}_{\psi}=1$ and $y^{c,u}_{\psi}=0.01,0.5$. However, since the coupling to up quarks, $y^{u}_{\psi}$, is larger, collider constraints coming from monojet searches begin to probe the low $m_{\psi}$ region.
Last but not least, in Figure~\ref{fig:Results_G3}, we show our results for the benchmark $y^{t}_{\psi}=0.4$ and $y^{c,u}_{\psi}=0.01,1$. Here the collider constraints are purely dominated by monojet searches for low Majorana neutrino masses and by jets + MET searches in the large $M_{N_{R}}$ region, since in the latter the production is dominated by the diagram depicted in Figure~\ref{fig:multijet2}$(b)$. In addition, since the value of $y^{t}_{\psi}$ is smaller that in the previous two benchmarks, a region of parameter space is excluded by $\mu\to e\gamma$. This is consistent with the fact that the branching ratio for this decay is inversely proportional to $K^{t,t}$ already for values of $y^{c}_{\psi}$ of order $10^{-2}$.  This region is depicted below the dashed black line.

We see that for the scenarios depicted in Figures~\ref{fig:Results_G1} and~\ref{fig:Results_G2}, a dark matter candidate with a mass of $200$ GeV is allowed by all experimental constraints for coloured electroweak-singlet scalars with masses between $400-600$ GeV. This situation can be relaxed if one sets the $y^{c}_{\psi}$ coupling to zero, completely eliminating the constraint form $D^{0}-\bar{D}^{0}$ oscillations. For this choice of parameters, a Majorana Neutrino mass below $100$ GeV is viable for $y^{u}_{\psi}$ couplings as large as $0.5$. This is a promising scenario since in this region of parameter space the top coupling, $y^{t}_{\psi}$ is also large and a monotop signature may be probed with the current data set in the semileptonic decay mode of the top quark and in the future $14$ TeV run.

\section{Monotop production}\label{sec:monotop}
In this section we discuss the monotop signal within our framework. The CDF collaboration reported on a search for dark matter using  $7.7$ fb$^{-1}$ of integrated luminosity and $p\bar{p}$ collisions at $\sqrt{s}=1.96$ TeV~\cite{Aaltonen:2012ek}. In addition, a stronger bound was set by the CMS collaboration using $19.7$ fb$^{-1}$ of integrated luminosity and $pp$ collisions at $\sqrt{s}=8$ TeV~\cite{CMS:2014hba}. Within the framework of effective field theories, they were able to place an upper bound on the production cross section of a dark matter particle in association with a single top quark. Their reach extends to scalar and vectorial dark matter masses below $\sim 327$ GeV and $655$ GeV respectively.
\begin{figure}[!ht]
\begin{center}\begin{picture}(230,100)(0,-60)
\small

\Gluon(55,10)(95,-21){4}{5}
\Vertex(95,-21){1.5}
\ArrowLine(55,-48)(95,-21) \Text (65,-50)[]{$c/u$}

\ArrowLine(95,-21)(130,-21) \Text (115,-13)[]{$c/u$}

\Vertex(130,-21){1.5}
\Line(130,-21)(170,10) \Text (150,10)[]{$N_{R}$}
\DashLine(130,-21)(170,-48){5} \Text (145,-50)[]{$\psi$}

\Vertex(170,-48){1.5}
\ArrowLine(170,-48)(220,-37) \Text (225,-37)[]{$t$}
\Line(170,-48)(220,-59) \Text (230,-59)[]{$N_{R}$}

\end{picture}
\end{center}
\caption{\small Leading order Feynman diagram for monotop production at the LHC in association with missing transversed energy carried away by Majorana neutrino.}
\label{fig:monotop}
\end{figure}
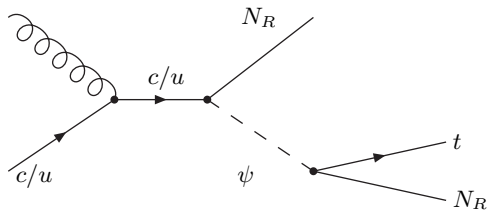

 In order to compliment the above searches as well as other models that predict a monotop signal, we simulate our signal with an additional jet, that is, contributions to the rate arise from the diagram depicted in Figure~\ref{fig:monotop} as well as Figures~\ref{fig:multijet1}($a$) and~\ref{fig:multijet2}($a$) where one coloured electroweak-singlet decays to a top quark while the other to a light jet. We implement a search strategy at the LHC in the semileptonic decay mode of the top quark, that is: $pp\to t+N_{R}N_{R}\to bl\nu+N_{R}N_{R}$. We study the possibility of probing the allowed region of parameter space with a monotop signal using the full data set with $8$ TeV centre of mass energy as well as in the future $14$ TeV run.

\subsection{LHC at $8$ TeV: Semileptonic mode.}

In order to probe a monotop signal at the LHC in the semileptonic decay mode of the top quark one must overcome the very challenging feat of reducing the large QCD multijet and the $t\bar{t}$ backgrounds. In what follows we implement a monotop search strategy developed in~\cite{Alvarez:2013jqa}, where one tags the top quark through its semileptonic decay mode. The authors show that the key kinematic variable used to suppress most of the SM backgrounds is the transverse mass of the charged lepton, $M_{T}$, given by
\begin{equation}
M_{T}=\sqrt{2p^{l}_{T}\slashed{E}_{T}(1-\cos\Delta\phi_{l,\slashed{E}_{T}})},
\end{equation}
where $p^{l}_{T}$ is the transverse momentum of the lepton arising from the decay of the top quark, and $\Delta\phi_{l,\slashed{E}_{T}}$ is the angular separation between the lepton and the missing energy. For SM backgrounds, this variable tends to peak for smaller values since a real $W$ can be reconstructed. This is not the case for the signal since additional sources of missing energy arise from the Majorana neutrino, $N_{R}$.

The dominant backgrounds are simulated at leading order using MadGraph 5~\cite{Alwall:2011uj}. We implement PYTHIA~\cite{Sjostrand:2006za} for the parton showering and hadronization. We use the MLM matching scheme~\cite{Mangano:2006rw} to avoid double counting. The detector simulation is carried out using Delphes 3~\cite{delphes}, and it is used for jet clustering and lepton isolation. In addition, we implement a $b$-tagging efficiency of $40\%$, a charm misidentification probability of $10\%$ and $0.1\%$ for light jets. This requirements are consistent with the PGS $b$-tagging efficiencies for tight tags~\cite{PGS}. We then reweight  the events to include higher order corrections when they are available. A $k$ factor for $t\bar{t}$ production is obtained by normalizing the leading order (LO) inclusive production cross section to the next-to-next-to-leading order (NNLO) cross section calculated in~\cite{NNLOttbar}. For single top production we simulate $pp\to tj$, $pp\to\bar{t}j$, $pp\to tW$ and $pp\to\bar{t}W$ and normalize the total cross section to the next-to-next-to-leading log threshold resummed result in~\cite{NNLLt}. For $W$ and $Z$ production we simulate the LO inclusive cross section and decay the gauge bosons leptonically. We then normalize the $pp\to W~X\to l\nu~X$ cross section to the NNLO result in~\cite{NNLOwX} and the $pp\to Z~X\to l^{+}l^{-}~X$ in~\cite{NNLOzX}. The LO order diboson production are normalized to their next-to-leading order (NLO) predictions in~\cite{NLOdiboson}. The $k$-factors associated with the dominant backgrounds are summarized in Table~\ref{tab:kfac}.
\begin{table}[ht]\centering
\begin{tabular}{|c|c|c|c|c|c|c|c| }
\hline
& $W~X,W\to l\nu_{l}$ &  $Z~X,Z\to ll$ & $t\bar{t}$ & $tj+tW$ & $WZ$ & $ZZ$ & $WW$\\
\hline
\hline
LO $\sigma$ & 10.78 nb & 0.9416 nb & 191.7 pb & 100.16 pb & 12.92 pb & 4.90 pb & 34.8 pb \\
\hline
NNLO $\sigma$ & 12.50 nb & 1.13 nb & 245.8 pb & 114.95 pb & 22.87 pb & 7.94 pb  & 57.42 \\
\hline
$k$-factor & 1.16 & 1.20 & 1.28 & 1.15 & 1.77 & 1.62 & 1.65 \\
\hline
\end{tabular}
\caption{\small $k$-factors for the leading SM backgrounds. The various next-to-leading results are summarized in the text. } \label{tab:kfac}
\end{table}

We simulate the $t\bar{t}$ background with up two jets. This background can be suppressed by demanding one isolated lepton and requiring $M_{T}\gtrsim 80$ GeV, since we expect most of the missing energy to come from a reconstructed $W$. Additional missing energy may arise from misreconstructed jets but these events can be suppressed by vetoing on jets with large $p_{T}$. Single top production is simulated with up to one jet or a $W$ boson. This background is irreducible when the production is in association with a jet and can produce additional missing energy in the $tW$ mode in case a lepton is missed. We simulate $Wj$ with up to three jets. This background has a large cross section, but vetoing on events with more than one jet and requiring that $M_{T}\gtrsim 80$ GeV  significantly reduces the background. $Zj$ is also simulated with up to three jets, and can be suppressed by requiring one isolated lepton and vetoing on events with more than one jet. Furthermore, the contributions from $Zj/Wj$ are suppressed with the $b$-tagging requirements described above. We simulate $WW$, $WZ$ and $ZZ$, but note that this background contributes the most if a jet is mistagged as a $b$-jet. The QCD multijet background comes from misidentified leptons and the missing energy from misreconstructed jets. The authors in~\cite{Alvarez:2013jqa} argue that a high $p_{T}$ veto is very effective at suppressing the QCD multijet background that we do not simulate and it is efficient at also reducing the all hadronic decay mode the $t\bar{t}$ background.

\begin{figure}[ht]\centering
\subfigure{\includegraphics[width=3.0in]{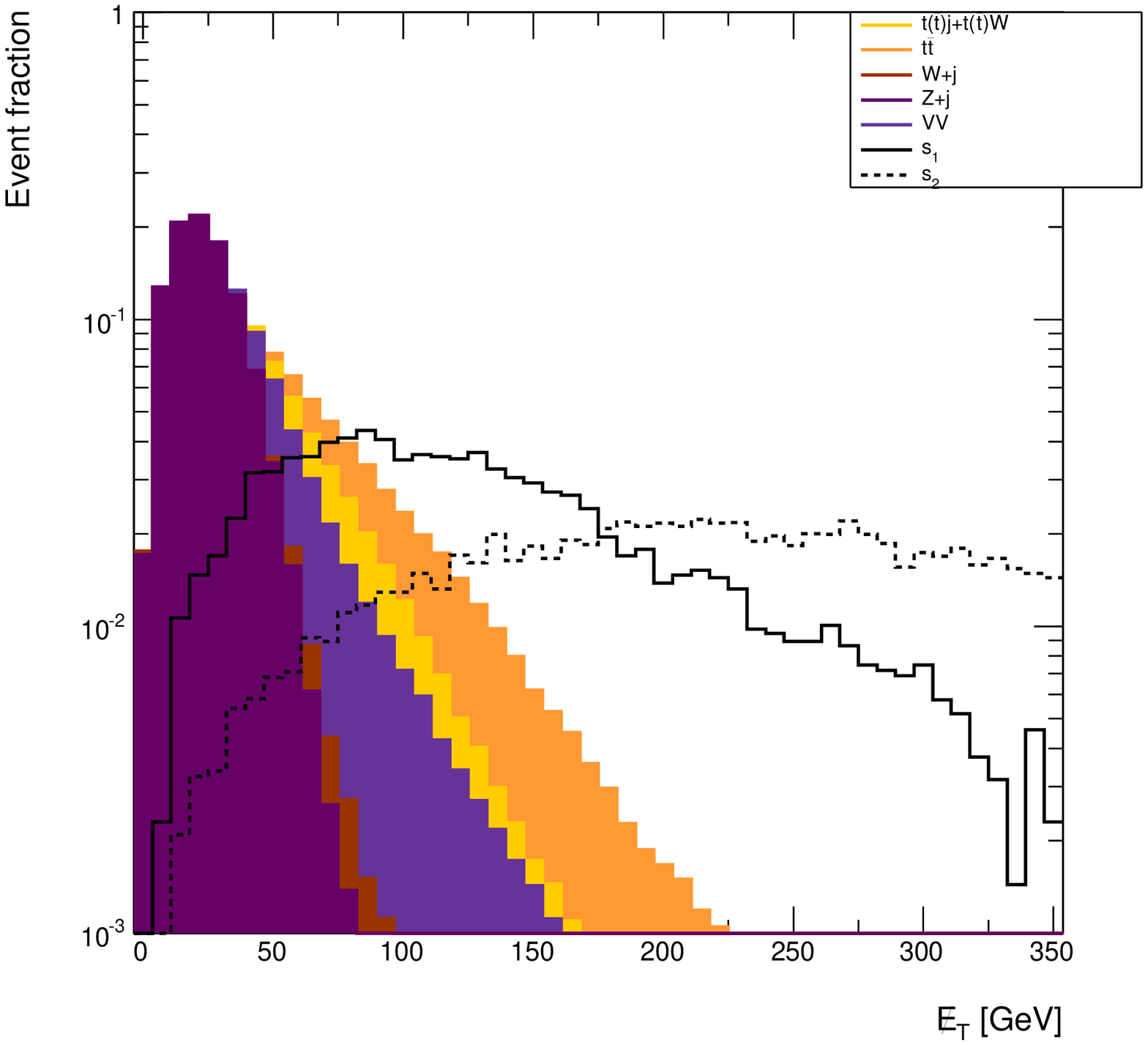}\label{fig:met_G2}}
\subfigure{\includegraphics[width=3.0in]{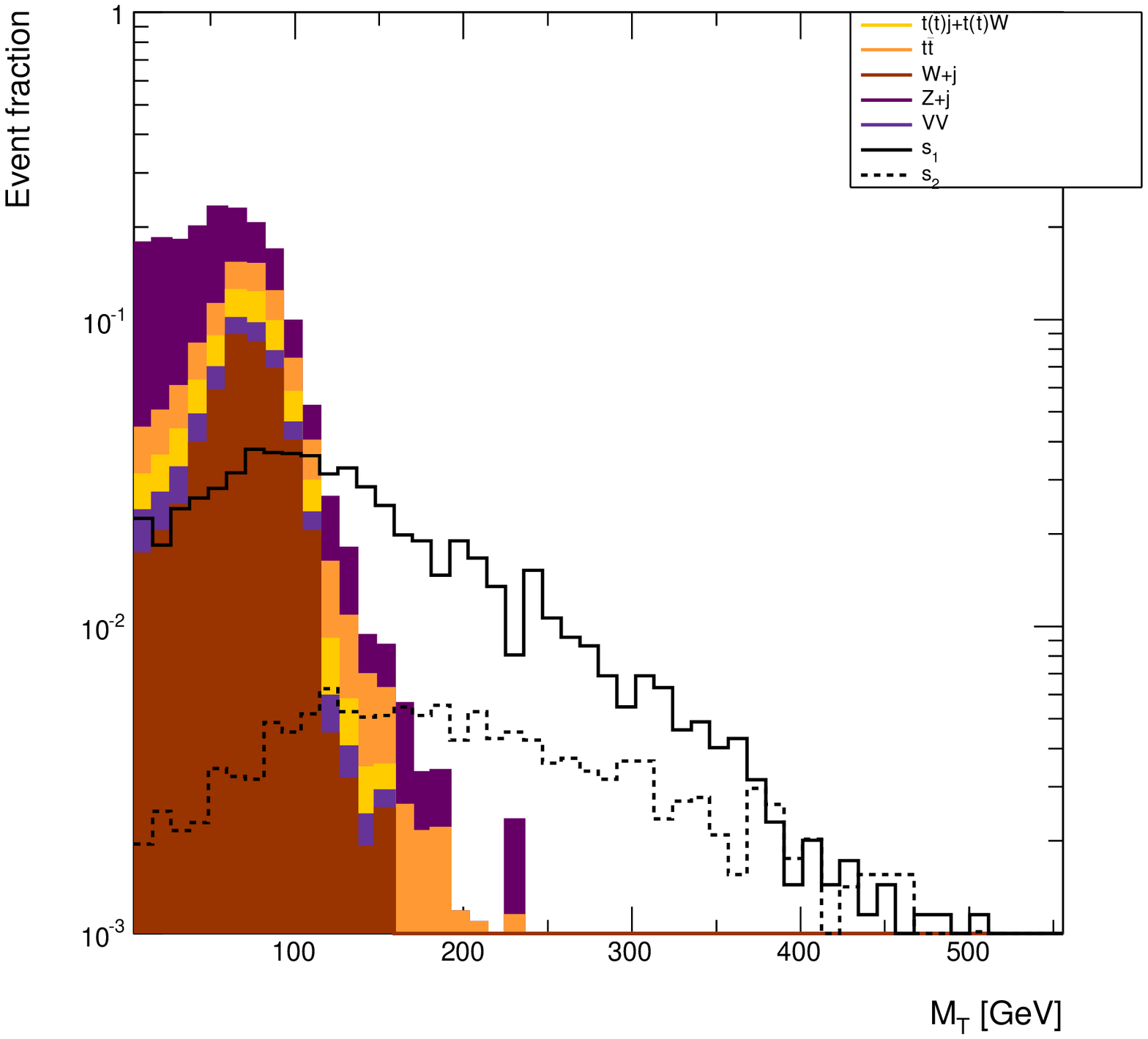}\label{fig:mtlep_G2}}
\caption{\small \small Fraction of events as a function of the missing transverse energy, $\slashed{E}$ (left) and the lepton's transverse mass, $M_{T}$ (right). The black solid line represents a scenario within our framework where $m_{\psi}=150$ GeV and $M_{N_{R}}=80$ GeV while the dashed black line represents $m_{\psi}=700$ GeV and $M_{N_{R}}=210$ GeV. Both signals were generated using $(y^{t}_{\psi},y^{c}_{\psi},y^{u}_{\psi})=(1,0,0.5)$.  }\label{fig:G2LepMain}
\end{figure}

We pre-select events by requiring one charged lepton with $p_{T}>20$ GeV and $|\eta|<2.5$ in addition to the presence of a $b$-jet with $p_{T}>20$ GeV and $|\eta|<2.5$. We require one light jet with $p_{T}>20$ GeV and $|\eta|<4.5$. The effects of pile up on the missing energy, in addition to those implemented in Delphes, are taken into account using a gaussian function to smear the amount of missing energy based on the total $p_{T}$ in the event~\cite{AtlasSmear}. In Figures~\ref{fig:met_G2} and~\ref{fig:mtlep_G2} we show the fraction of events as a function of MET and the transverse mass, $M_{T}$, only after requiring one $b$-tagged jet. The solid black line corresponds to a low mass scenario where $M_{N_{R}}=80$ GeV and $m_{\psi}=150$ GeV and the dashed black line to a high mass scenario where  $M_{N_{R}}=210$ GeV and $m_{\psi}=700$ GeV. This is done in order to identify two signal regions for all of the parameter space parametrized in the $m_{\psi}-M_{N_{R}}$ plane. Using these two distributions we define a low mass signal region where $\slashed{E}_{T}>90$ GeV and $M_{T}>110$ GeV, and a high mass signal region where $\slashed{E}_{T}>200$ GeV and $M_{T}>120$ GeV.  Furthermore, in order to suppress the QCD multijet background, in samples with one light jet, we exclude any event where $p_{T,j}>70,120$ GeV for the two sets of cuts respectively since misreconstructed jets may appear as large missing energy. In Table~\ref{tab:cuts_lep} we show the number of expected events and cross section for the SM backgrounds for the two signal regions. One can see that the $Z+$jets background is absent in both cases, while a high MET cut and a large jet $p_{T}$ veto significantly reduces the $t\bar{t}$ and $W+$jets backgrounds.
\begin{table}[ht]\centering
\begin{tabular}{|c|c|c| }
\hline
SM background & $N^{\slashed{E}>90~\text{GeV},M_{T}>110~\text{GeV},p_{T,j}<70~\text{GeV}}$ ($\sigma$ [pb]) & $N^{\slashed{E}>200~\text{GeV},M_{T}>120~\text{GeV},p_{T,j}<120~\text{GeV}}$ ($\sigma$ [pb])  \\
\hline
\hline
$W+$ jets             &   $1375\pm37$ ($6.87\times10^{-2}$)                                                          & $\approx 0$ \\
\hline
$t\bar{t}+$ jets     &   $3104\pm56$ ($1.55\times10^{-1}$)                                                          &  $74\pm9$ ($3.71\times10^{-3}$) \\
\hline
$t~j+t~W$            &   $662\pm26$ ($3.31\times10^{-2}$)                                                          &  $10\pm3$ ($5.06\times10^{-4}$) \\
\hline
$WW$                  &     $16\pm4$ ($8.02\times10^{-4}$)                                   & $1\pm1$ ($5.73\times10^{-5}$) \\
\hline
$WZ$                     &    $11\pm3$ ($5.71\times10^{-4}$)                                  & $\approx 0$ \\
\hline
$ZZ$                     &     $2\pm1$ ($7.97\times10^{-5}$)                                   & $\approx 0$   \\
\hline
\end{tabular}
\caption{\small Number of expected events, $N$, with $20$ fb$^{-1}$ of integrated luminosity and cross section in the two signal regions specified by the amount of MET ($>90$, $>200$ GeV) and charged lepton's transverse mass, $M_{T}~(>110,120~\text{GeV})$} for the dominant SM backgrounds. \label{tab:cuts_lep}
\end{table}

\begin{figure}[ht]\centering
\subfigure{\includegraphics[width=3.0in]{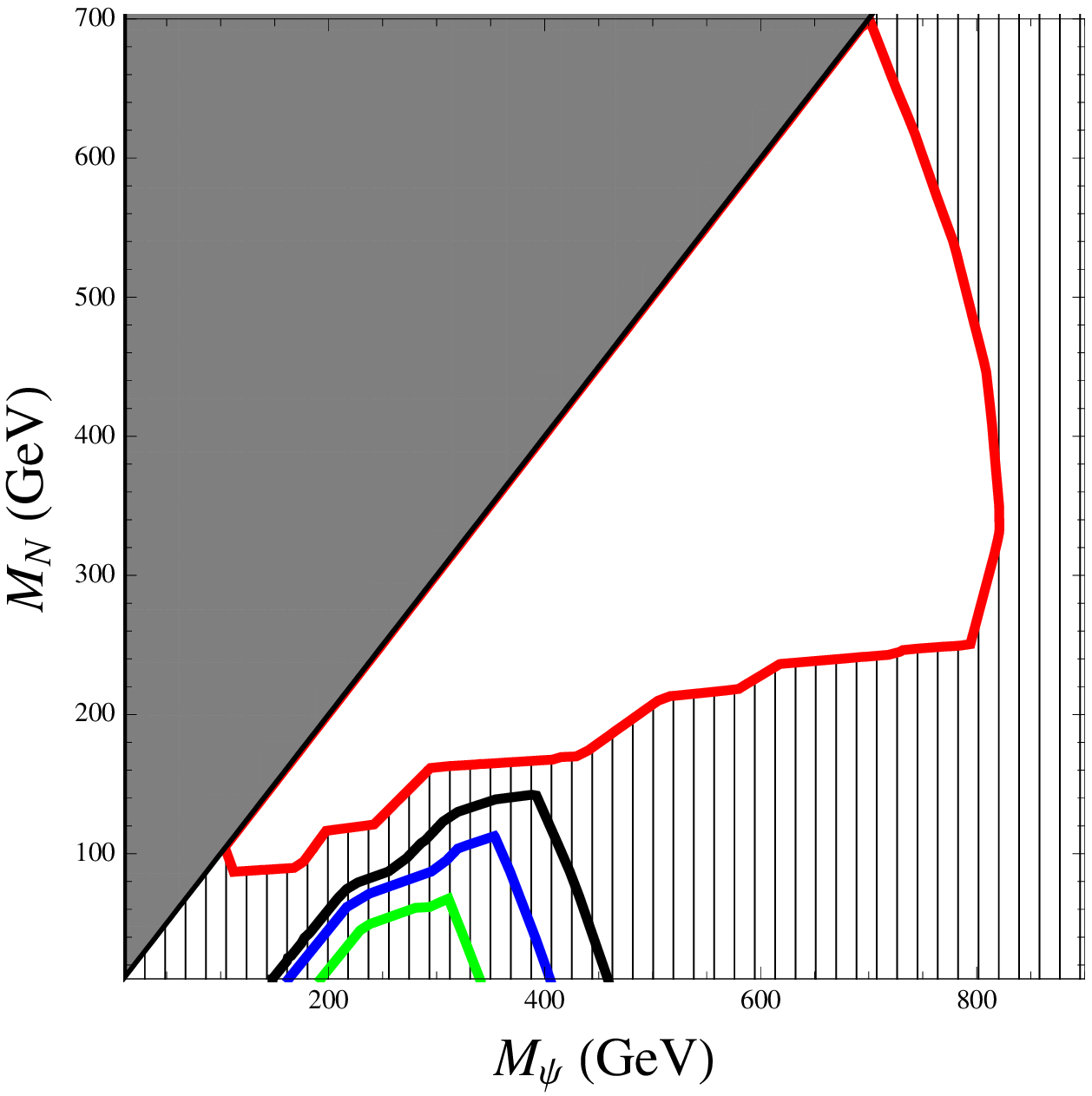}\label{fig:monotop_LEP_LM_G2}}
\subfigure{\includegraphics[width=3.0in]{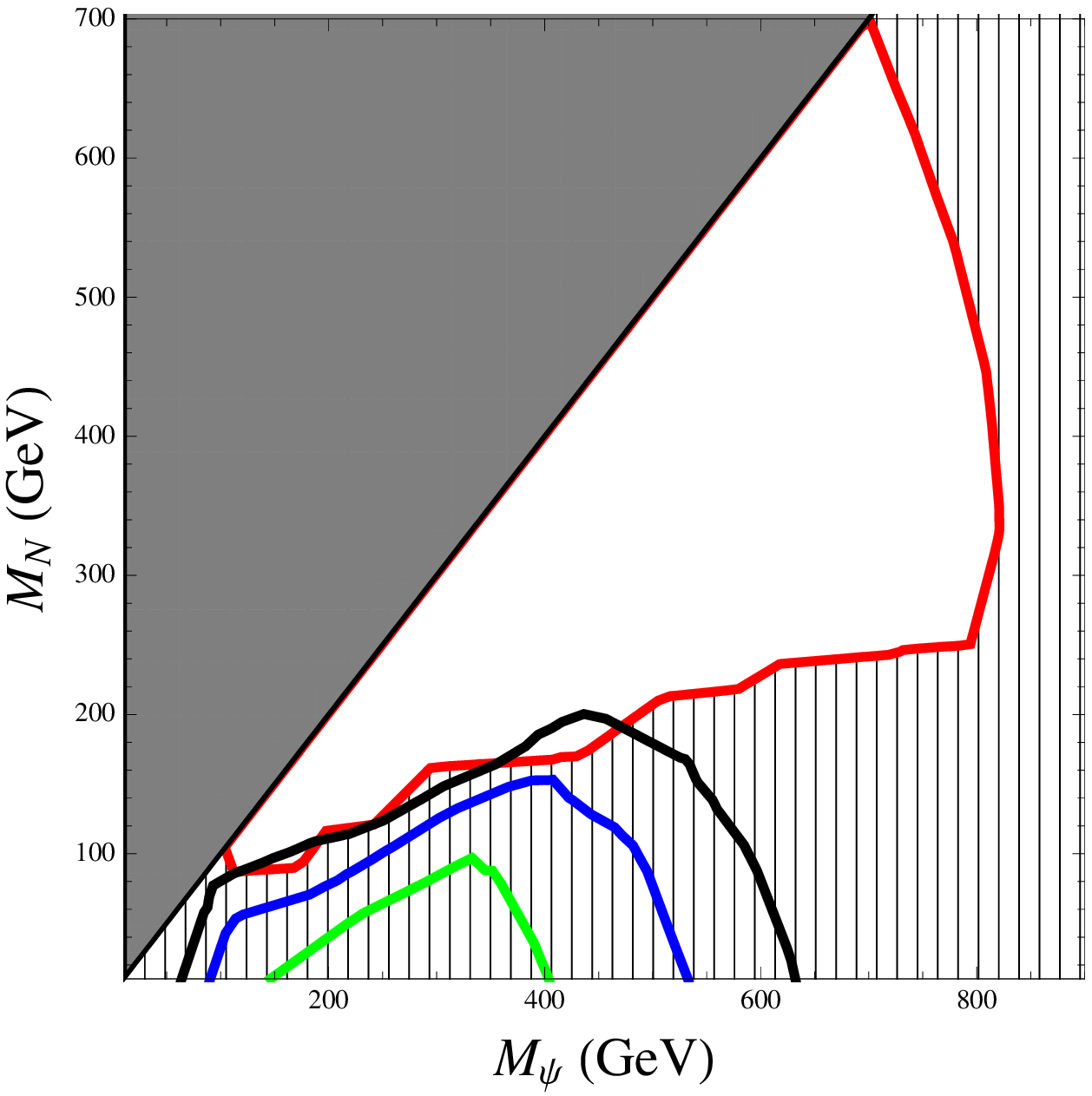}\label{fig:monotop_LEP_HM_G2}}
\caption{\small LHC reach with 20 fb$^{-1}$ at $8$ TeV of our semileptonic monotop signal using $(y^{t}_{\psi},y^{c}_{\psi},y^{u}_{\psi})=(1,0,0.5)$ after applying all the cuts described in the text with $\slashed{E}_{T},M_{T}>90,110$ GeV (left) and $\slashed{E}_{T},M_{T}>200,120$ GeV (right). We show three regions where our monotop signal can reach a significance, $s=S/\sqrt{S+B}$, of two, three and five depicted by the dashed black, blue and green lines respectively. }\label{fig:G2LepMain}
\end{figure}

In Figures~\ref{fig:monotop_LEP_LM_G2} and~\ref{fig:monotop_LEP_HM_G2} we show the allowed region of parameter space after all constraints have been taken into consideration (light and dark grey regions) in the $m_{\psi}-M_{N_{R}}$ plane for the two signal regions using as a benchmark the couplings $(y^{t}_{\psi},y^{c}_{\psi},y^{u}_{\psi})=(1,0,0.5)$. We saw in the previous section that this benchmark was the most promising for a monotop signal after taking into account all constraints. The region excluded by all of the constraints described earlier (using CMS data for the direct collider searches) is denoted by the vertical hatched region, the white region is an allowed region that does not account for $100\%$ of the observed relic abundance and the solid red line yields a thermal relic. The black, blue, and green solid lines depict the region of parameter space where our monotop signal can reach a significance, $s=S/\sqrt{S+B}$, of two, three and five respectively. One can see that cuts consistent with the high mass signal region are more efficient at eliminating the SM backgrounds, and this model can be probed at the two sigma level for $m_{\psi}$ between $100$ and $500$ GeV and $M_{N_{R}}$ between $100$ and $200$ GeV.

\subsection{LHC at $14$ TeV}
In this section we analyze the LHC reach of our monotop signal at $14$ TeV centre of mass energies with $30$ and $300$ fb$^{-1}$ of integrated luminosities. We give an estimate of the SM backgrounds using the $k$-factors introduced in Table~\ref{tab:kfac} to account for the higher order QCD corrections and simulate the cross sections at leading order using Madgraph 5~\cite{Alwall:2011uj}. We implement PYTHIA~\cite{Sjostrand:2006za} for the parton showering and hadronization. We use the MLM matching scheme~\cite{Mangano:2006rw} to avoid double counting. The detector simulation is carried out using Delphes 3~\cite{delphes}, and it is used for jet clustering and lepton isolation. In addition, we implement a $b$-tagging efficiency of $40\%$, a charm misidentification probability of $10\%$ and $0.1\%$ for light jets. The normalized cross sections are given in Table~\ref{tab:presel_14tev}. In the previous section we saw that one may probe this model with $20$ fb$^{-1}$ of data at $\sqrt{s}=8$ TeV with a significance, $s$, of two. Thus, probing this model with future LHC energies is important in order to explore the remaining region of parameter space. We focus again on the benchmark point given by $(y^{t}_{\psi},y^{c}_{\psi},y^{u}_{\psi})=(1,0,0.5)$.

\begin{figure}[ht]\centering
\subfigure{\includegraphics[width=3.0in]{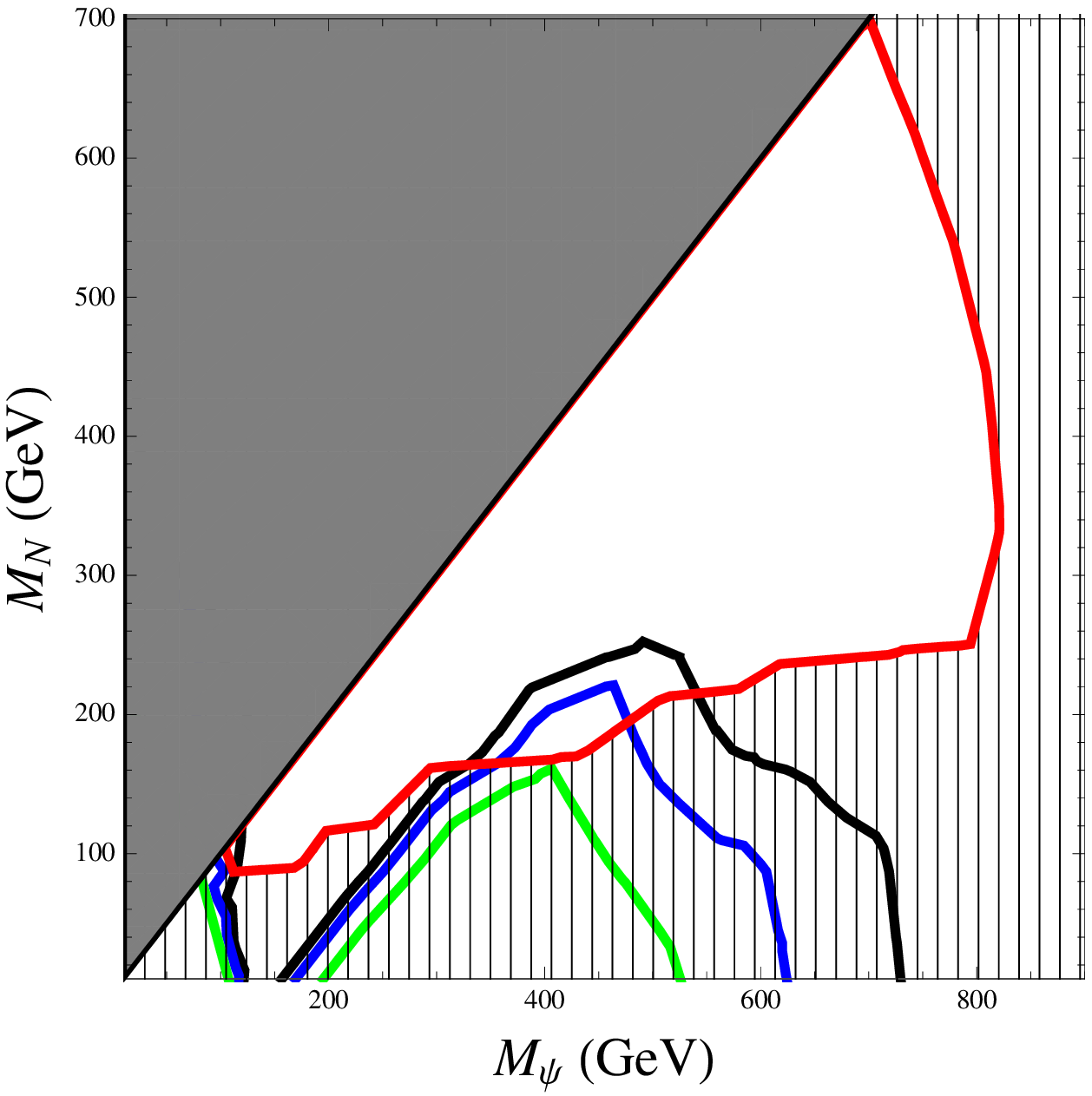}\label{fig:monotop_LEP_HM_G2_14tev_30fb}}
\subfigure{\includegraphics[width=3.0in]{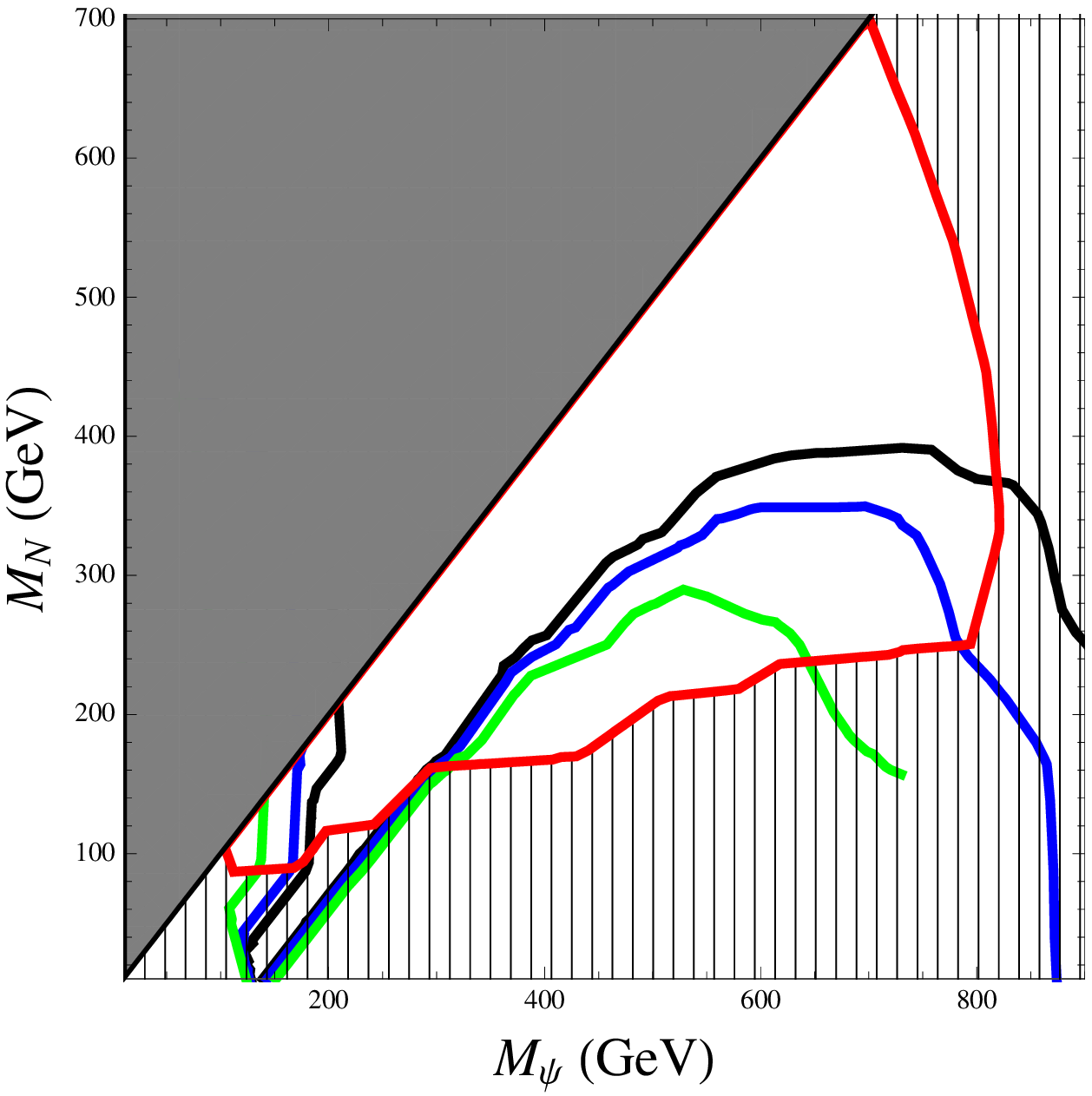}\label{fig:monotop_LEP_HM_G2_14tev_300fb}}
\caption{\small LHC reach with $30$ (left) fb$^{-1}$ and $300$ fb$^{-1}$ (right) at $14$ TeV of our semileptonic monotop signal using $(y^{t}_{\psi},y^{c}_{\psi},y^{u}_{\psi})=(1,0,0.5)$ and after applying all the cuts described in the text. The three regions where our monotop signal can reach a significance, $s=S/\sqrt{S+B}$, of two, three and five are depicted by the dashed black, blue and green lines respectively.}\label{fig:G2LepMain14}
\end{figure}

\begin{table}[ht]\centering
\begin{tabular}{|c|c| }
\hline
Process & $\sigma$ [pb] \\
\hline
\hline
$W+$ jets & $2.19\times10^{5}$ \\
\hline
$Z+$ jets & $6.66\times10^{4}$ \\
\hline
$t\bar{t}+$ jets & 1052.93 \\
\hline
$tj$+$tW$ & 347.42 \\
\hline
$WW$ &      119.84          \\
\hline
$WZ$ &       48.87        \\
\hline
$ZZ$ &          17.09      \\
\hline
\end{tabular}
\caption{\small Cross sections normalized to their next-to-leading order results for the leading SM background contributions at centre of mass energies of $14$ TeV. The cross sections are normalized using the $k$-factors introduced in Table~\ref{tab:kfac}. } \label{tab:presel_14tev}
\end{table}
In the previous section, we showed that the full data set at $\sqrt{s}=8$ TeV was not enough to probe the parameter space using the semieptonic decay mode of the top quark. However, we saw that a large enough cut on the MET was enough to suppress the $W$ and $Z$ plus jets backgrounds. This was due to the low acceptance rate from demanding a $b$-jet and an isolated lepton in the final state. The suppression of the $W$ plus jets background component was also due to a cut on the transverse mass of the charged lepton, $M_{T}$. At energies of $\sqrt{s}=14$ TeV, the same behaviour is observed if one keeps events with a high enough MET. Therefore, in our analysis we apply the cuts used to enhance the sensitivity of the search to the high $m_{\psi}$ region:
\begin{equation}
\slashed{E}_{T}>200~\text{GeV},~~~M_{T}>120~\text{GeV}.
\end{equation}
In addition, we require events to contain up to one jet with $p_{T}<120$ GeV in order to suppress the QCD multijet SM background which we do not simulate. The signal cross section together with the three dominant backgrounds, $Wj,t\bar{t}$ and $tj+tW$, are shown in Table~\ref{tab:cuts_lep_14}. The signal corresponds to a coloured electroweak-singlet scalar with mass $m_{\psi}=700$ GeV and a Majorana neutrino with mass $M_{N_{R}}=210$ GeV. In Figures~\ref{fig:monotop_LEP_HM_G2_14tev_30fb} and~\ref{fig:monotop_LEP_HM_G2_14tev_300fb} we show the signal significance in the $m_{\psi}-M_{N_{R}}$ plane using $30$ fb$^{-1}$ and $300$ fb$^{-1}$ of integrated luminosities respectively.  Contours of $s=2,3,5$ are depicted by the solid black, blue and green lines respectively. Compared to the scenario depicted in Figure~\ref{fig:monotop_LEP_HM_G2}, the increase in energy from $8$ to $14$ TeV already shows that the LHC will potentially begin to probe this framework very early during run 2. However, an order of magnitude increase in the luminosity will probe the allowed region of parameter space for Majorana neutrino masses up to $\sim400$ GeV, complimenting searches of pair production of scalar top quarks at the LHC. We can also see from the figure that at these luminosities one can begin to probe the compressed region where $m_{\psi}\approx m_{t}$.
\begin{table}[ht]\centering
\begin{tabular}{|c|c|c|c|c| }
\hline
${\cal L}$ [fb$^{-1}$] & $\sigma\left(W+\text{jets}\right)$ [pb], N  &       $\sigma\left(t\bar{t}+\text{jets}\right)$ [pb], N         & $\sigma\left(tj+tW\right)$ [pb], N  & $\sigma_{signal}$[pb], N \\
\hline
\hline
 $30$   &    $1.44\times10^{-2}$, $431\pm21$       &   $2.98\times10^{-2}$, $895\pm30$               &              $3.82\times10^{-3}$, $115\pm11$                                         & $2.00\times10^{-3}$, $60\pm8$ \\
\hline
$300$    &   $$ $4308\pm66$      &   $$ $8945\pm95$               &              $$ $1146\pm34$                                         & $600\pm24$ \\
\hline
\end{tabular}
\caption{\small Number of expected events, $N$, with $30$ and $300$ fb$^{-1}$ of integrated luminosities and cross section at $\sqrt{s}=14$ TeV for the three dominant SM backgrounds specified after applying the cuts mentioned in the text. The signal strength is also shown for $m_{\psi}=700$ GeV and $M_{N_{R}}=210$ GeV.} \label{tab:cuts_lep_14}
\end{table}

\begin{figure}[ht]\centering
\includegraphics[width=3.0in]{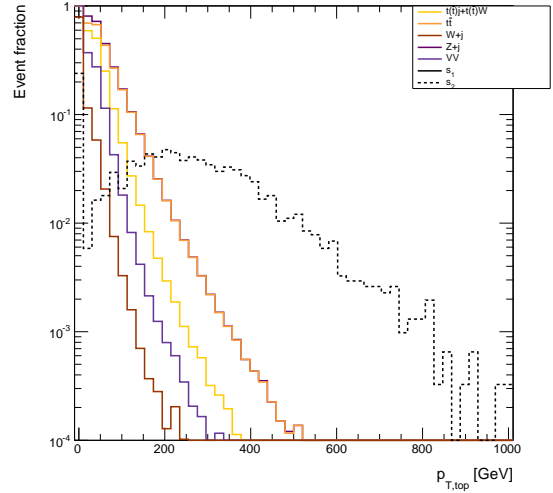}
\caption{\small Fraction of events as a function of the transverse momentum of the reconstructed top quark, $p_{T,top}$. The dashed black line represents a scenario within our framework where $m_{\psi}=700$ GeV and $M_{N_{R}}=210$ GeV. Both signals were generated using $(y^{t}_{\psi},y^{c}_{\psi},y^{u}_{\psi})=(1,0,0.5)$.  }\label{fig:ptT_14tev_G2}
\end{figure}
Even though the above analysis has been carried out by tagging the semileptonic decay mode of the top quark, one can equally tag the hadronic decay mode. However, the all-hadronic and QCD multijet backgrounds do represent a problem in accurately taking into account all SM backgrounds. Simulating these backgrounds will require the combined efforts from experimentalists and theorists alike. Nonetheless, one can start to better discriminate the signal from the well established SM backgrounds by tagging boosted tops. Within our framework, at $\sqrt{s}=14$ TeV, the production of heavy coloured electroweak-singlet scalars leads to boosted tops provided that the mass of the Majorana neutrino is not very large. This is particularly interesting since this region of parameter space is consistent with the relic abundance of dark matter in the universe. In Figure~\ref{fig:ptT_14tev_G2} we show the fraction of events as a function of the transverse momentum of the reconstructed three-jet system which we label $p_{T,top}$ only after requiring that each event contains a $b$-tagged jet. The momentum corresponds to the combination of the leading $b$-jet in the event together with the leading two light jets. For the signal (black dashed line) this is consistent with the transverse momentum of a top quark. We see that a larger fraction of the events peaks at values greater than $200$ GeV and the presence of boosted top quarks, unlike the signal where most of the SM background has a sharp drop. The study of boosted tops and how they are tagged is an active field of research~\cite{Thaler:2008ju,Cambridge,Kaplan:2008ie,Plehn:2010st,Thaler:2010tr,Thaler:2011gf} and it would be very interesting to see its effects on models that predict a monotop signal at large centre of mass energies, together with a full implementation and treatment of the QCD multijet background.

\section{Discussion}\label{sec:discussion}

In this study we have generalized the model introduced in~\cite{ModelMain}, coupling all three generations of right-handed up-type quarks to a Majorana neutrino and a coloured electroweak-singlet scalar. Within this framework, the dark matter relic abundance can match the latest experimental results over a wide range of couplings $y^{t,c,u}_{\psi}$. However, we saw that very large values of $y^{u}_{\psi}$ can be in disagreement with limits obtained from direct dark matter searches. In particular, the model is mainly constrained by spin dependent interactions between the dark matter and nuclei. The constraints are larger when the coloured electroweak-singlet scalar is in resonance with the Majorana neutrino. Furthermore, the possibility of radiatively generating Majorana masses for the active neutrinos is mainly dependent on $y^{t}_{\psi}$ since the contributions from the up and charm quarks are proportional to their masses. Natural neutrino masses are possible over the whole range of $y^{t}_{\psi}$ values. However, constraints from rare decays such as $\mu\to e\gamma$ prefer large values of $y^{t}_{\psi}$. In addition, $D$-meson oscillations tend to constrain the product $y^{u}_{\psi}y^{c}_{\psi}$.

We have implemented collider searches for monojets and multijets in association with missing transverse energy. In addition, it was seen that searches for scalar top pair production can further exclude a wide region of the parameter space. To analyze the impact of these searches, we have applied the constraints to three benchmark scenarios. The benchmark scenarios all have a large value of $y^{t}_{\psi}$ to evade rare decay bounds and small value of $y^{c}_{\psi}$ to avoid constraints from flavour oscillations. Three values of $y^{u}_{\psi}$ ranging from $0.1-1$ were considered. This coupling was varied mainly because we study the production of a single top quark in association with missing transverse energy, which in this framework, has as the main production mode quark-gluon fusion. This production mode is enhanced for large values of $y^{u}_{\psi}$, however, the parameter region corresponding to large $y^{u}_{\psi}$ is highly constrained by monojet~+~MET searches.

We have analyzed the monotop production in the semileptonic decay mode of the top quark after implementing a search by the CMS collaboration in the hadronic decay mode at $\sqrt{s}=8$ TeV. It was seen that with the current $\sqrt{s}=8$ TeV data set a monotop search does not probe the allowed region of parameter space. The situation changes at $\sqrt{s}=14$ TeV, where an approximate calculation puts this model within the reach of the LHC with $30$ and $300$ fb$^{-1}$ of integrated luminosity. Future work may want to consider better top quark reconstruction techniques to better discriminate the SM background. In particular, tagging boosted tops can be used to better probe models that predict a monotop signature with $14$ TeV centre of mass energies.

\section*{Acknowledgements}

The authors would like to thank Estefania Coluccio Leskow, Travis A.W. Martin, Kristian L. McDonald, and David Morrissey for useful discussions and essential feedback regarding the progress of this work. This work is supported in parts by the National Science and Engineering Council of Canada.


\end{document}